\documentclass[preprint,12pt]{elsarticle}
\usepackage{amsmath}
\usepackage{graphicx}% Include figure files
\usepackage{dcolumn}% Align table columns on decimal point
\usepackage{bm}% bold math
\usepackage{caption}
\usepackage{float}
\usepackage{adjustbox}
\usepackage{xcolor}
 \RequirePackage{mathptmx} 
\journal{Nuclear Physics A}

\begin{document}

\begin{frontmatter}

\title{Re-examination of nuclear structure properties and shape co-existence of nuclei around A $\sim$ 70}
\author{{Jameel-Un Nabi$^{1}$, Tuncay Bayram$^{2}$, Mahmut B\"{o}y\"{u}kata$^{3}$, Asim Ullah$^{4}$\footnote{asimullah844@gmail.com}, Anes Hayder$^{2}$ and Syeda Zainab Naqvi$^1$}}

\address{$^1$University of Wah, Quaid Avenue, Wah Cantt 47040, Punjab, Pakistan}
\address{$^2$Department of Physics, Faculty of Science, Karadeniz Technical University, 61080, Trabzon, T\"{u}rkiye}
\address{$^3$ Department of Physics, {Faculty of Engineering and Natural Sciences}, K\i r\i kkale University, 71450, K\i r\i kkale, T\"{u}rkiye}
\address{$^{4}$Department of Physics, University of Swabi, Swabi 23561, KP, Pakistan}
 
                    %% ABSTRACTT %%   
\begin{abstract}
	We re-examine the  nuclear structure properties of waiting point nuclei around A $\sim$ 70 using the interacting boson model-1 (IBM-1) and the relativistic mean field (RMF) model. Effective
density-dependent meson-exchange functional (DD-ME2) and density-dependent point-coupling
functional (DD-PC1) were used for the RMF calculations. We calculated the energy levels, the geometric shapes, binding and separation energies of nucleons and quadrupole deformation parameters ($\beta_2$). The shape co-existence phenomena in  A $\sim$ 70 nuclei ($^{68}$Se, $^{70}$Se, $^{70}$Br, $^{70}$Kr, $^{72}$Kr, $^{74}$Kr, $^{74}$Rb and $^{74}$Sr) was later investigated. Spherical and deformed shapes of the selected waiting point nuclei were computed using the IBM-1 and RMF  models, respectively. The proton-neutron quasiparticle random phase approximation (pn-QRPA) model was used to calculate $\beta$-decay properties (Gamow-Teller strength distributions, $\beta$-decay half-lives and branching ratios) of selected nuclei  as a function of $\beta_{2}$.  The results revealed a significant variation in calculated half-lives and Gamow-Teller strength distributions as the shape parameter was changed. The $\beta_{2}$ computed via DD-ME2 functional resulted in half-lives in best agreement with the measured data.	
\end{abstract}
\end{frontmatter}
\section{Introduction}
\label{intro}
After Rutherford discovered the atomic nucleus, many exciting questions arose in the beginning of nuclear physics pertaining to the fundamental nature of the nuclear force and the intricate dynamics governing the coexistence of protons and neutrons in a very compact nucleus. 
%Once the existence of energy shells was revealed as the primary feature of nuclear structure by the groundbreaking work of Mayer \cite{Mayer}, these energy shells dictate the behavior of particle excitations and follow quantum rules defined by the nuclear Hamiltonian. The Shell Model, formulated shortly after Mayer's work, established a connection between collective nuclear phenomena and the individual behavior of protons and neutrons.
The geometric structure of the nucleus can provide information on the nature of the interactions between its constituents. The behavior of nuclear forces gives rise to intriguing phenomena, including shape co-existence. Low-lying 0$^+$ states in connection with 2$^+$ states above ground-state of even-even nuclei were deliberated much earlier~\cite{Boh53, Mor56}. Collective behavior of nucleon forces was held responsible for the deformation of the nucleus. Shape co-existence refers to the existence of multiple closely spaced eigenstates with varying intrinsic deformation within a finite nuclear many-body quantum system \cite{Hey11}. The shape-coexistence phenomena in atomic nuclei is a remarkable aspect of nuclear structure and the finite many-body quantum system. 
Shape coexistence allows researchers to examine how diverse sources contribute to correlation energy within the same nucleus, namely the quadrupole degree of freedom \cite{Pau22}. Shape transition and shape co-existence play a crucial role in comprehending the properties of low-lying nuclear structures and characterizing neutron-deficient isotopes \cite{Mor06, Bre14}. The phenomenon of shape co-existence is a manifestation of two opposing forces within the atomic nuclei. Whereas closed (sub)shells produce a stabilizing effect, the residual interactions between the nucleons tend to break the spherical symmetry and favor nuclear deformation.  The concept of shape co-existence \cite{Mor56} is observed in both light and heavy nuclei. Various reactions in the past (e.g., \cite{Mcl69}) provided evidence of shape co-existence and recent years have witnessed an increased emphasis on experimental as well as theoretical investigations in this field.\\
Neutron-deficient isotopes in the lead region serve as well-established examples defending shape co-existence phenomenon in nuclei \cite{Hey11}. Studies (e.g., \cite{Fri95,Sar98}) have revealed that the decay characteristics of $\beta$-unstable nuclei can be influenced by the nuclear shape of the decaying nucleus. Investigations have systematically examined the Gamow-Teller (GT) strength distributions associated with the $\beta^+$/EC-decay of neutron-deficient nuclei in the A $\sim$ 70 mass range \cite{Sar11}. These studies utilized a deformed quasiparticle random-phase approximation (QRPA) approach, with a self-consistent Hartree-Fock (HF) mean field and Skyrme forces, to analyze the dependence of GT strength on nuclear deformation. Similar analyses were later extended to stable $fp$-shell \cite{Sar13} and neutron-rich nuclei in the A $\sim$ 100 mass region \cite{Sar14}. The influence of deformation on GT strength distributions has proven to be a powerful tool in determining nuclear shape in neutron-deficient Kr and Sr isotopes. This was achieved by comparing theoretical predictions with  $\beta$-decay measurements~\cite{Poi13}.

The problem of shape co-existence becomes more complex for nuclei close to the N = Z line due to the competition between neutron-proton and like-nucleon interactions, which is expected to influence the behavior of these nuclei significantly \cite{Pet08}. Previously, the calculations of {the interacting boson model-1 (IBM-1) and the proton-neutron quasiparticle random phase approximation (pn-QRPA)} were performed to investigate the nuclear structure properties of the waiting point (WP) nuclei along N=Z chain around A $\sim$ 70 mass region ~\cite{Nabi16,Nabi17a}  {and $^{76}$Se isotope in same region ~\cite{Nabi17b}}. In the mass region A $\sim$ 70, few nuclei are predicted to possess a prolate surface state for N $\approx$ Z $\approx$ 38 \& 40 and an oblate surface state for N $\approx$ Z $\approx$ 36. Most of these nuclei are expected to showcase shape-mixing \cite{Pau22,Sar98,Dav21, Naz85,Bon85,Pet99,Pet15}. Consequently, numerous nuclear experiments have been conducted, and in some cases, strong predictions have been made regarding the prolate-oblate coexistence in this region. Neutron-deficient nuclei in the mass region A $\sim$ 70 are involved in rapid-proton capture ($rp$) nucleosynthesis and exhibit varying shapes and structural changes. These are attributed to the large gap in the spectrum of single particle energies at different deformations, which results in a robust struggle between varying many-body configurations based on corresponding deformations \cite{Pet00}. The astrophysical significance of the long-lived nuclei like $^{68-70}$Se and $^{72-74}$Kr could be that the proton capture increase the reaction flow, thus reducing the timescale for the $rp$-process nucleosynthesis during the cooling phase. The $\beta$-decay properties of the low-lying states in neutron-deficient nuclei, bearing relevance to $rp$-process, are argued to influence their effective $\beta$-decay half-lives for high temperatures prevailing in X-ray bursts~\cite{Sha98}.

%Stellar weak rate calculations have been performed for the waiting points of $^{68}$Se and $^{72}$Kr \cite{Pet15,Nabi16}.

%The structure of neutron-deficient nuclei $^{72}$Kr and $^{74}$Kr has been extensively studied in recent years, and an oblate-prolate competition strongly dominates at low spins. $^{72}$Kr exhibits an oblate shape, while $^{74}$Kr exhibits a prolate shape. However, the low- or high-spin states of $^{72}$Kr and $^{74}$Kr cores are still under investigation, posing challenges for microscopic nuclear models.

Nuclei in  A $\sim$ 70  region  are more dynamic in terms of their structural configurations, with diverse shapes coexisting and transitioning more readily than in other regions with equal numbers of neutrons or protons. We selected eight nuclei ($^{68}$Se, $^{70}$Se, $^{70}$Br, $^{70}$Kr, $^{72}$Kr, $^{74}$Kr, $^{74}$Rb and $^{74}$Sr) from this mass region for our investigation. The selected nuclei play crucial roles in the $rp$-process in astrophysical environments. They act as WP nuclei, influencing the reaction flow and nucleosynthesis calculations. The selected nuclei exhibit complex nuclear structures, including shape coexistence and collective properties, making them ideal candidates for studying the interplay of different nuclear shapes and configurations in the A $\sim$ 70 region. Advanced experimental techniques, such as decay studies and gamma-ray spectroscopy, enable researchers to investigate the decay properties and nuclear structure of the selected nuclei in detail.  Not all nuclei in the A $\sim$ 70 region are relevant to the $rp$-process or serve as WP nuclei. Only a subset of nuclei meet these criteria and were prioritized for the current study. We employ three different theoretical approaches in the current investigation: {the IBM-1, the relativistic mean field (RMF) and pn-QRPA model}. The IBM-1 and RMF models were used to calculate the nuclear structure properties. These include  energy levels, geometric shapes, binding and separation energies of nucleons and quadrupole deformation parameters ($\beta_2$) of the selected nuclei. The deformation parameters were calculated by plotting the potential energy surfaces (PESs). Later, the pn-QRPA model was used to study the $\beta$-decay properties of selected nuclei as a function of the computed deformation parameters. It is noted that the $\beta_2$ values enter as an input parameter in the pn-QRPA calculation.  Five different values of $\beta_2$, computed from the IBM-1 and RMF models,  were used as a free parameter in the pn-QRPA calculation.  A sixth value of deformation parameter was adopted from the FRDM calculation~\cite{Mol16}. The measured value of $\beta_2$, wherever available, was taken from the National Nuclear Data Center~\cite{NNDC23}.

The paper's structure is as follows: In Section 2, we briefly describe the theoretical framework employed in our calculations. Section 3 presents the outcomes and findings of the study. In Section 4, we provide a summary of our investigation along with some concluding remarks.

\section{Nuclear Models}

The necessary formalism of the three nuclear models, used in the current investigation, is described briefly in the succeeding subsections.

\subsection{Interacting Boson Model-1 (IBM-1)}
\label{ss_ibm}

The IBM-1 model proves to be effective in describing the nuclear structure properties of even-even nuclei~\cite{Iac87}.
This algebraic model is constructed on the six-dimensional $U(6)$ unitary group which has three possible subgroups called as \textit{dynamical symmetries}.
These symmetries are indicated by U(5), SU(3) and O(6). The value of energy ratio ($R_{4/2}$) of $4_1^+$ to $2_1^+$
levels in the ground state band can provide insight into
the geometrical behavior of the nucleus. The characteristic energy ratio values for SU(3), U(5), and O(6) symmetries are 3.33, 2.00, and 2.50 for axially deformed (oblate/prolate), spherical, and $\gamma$-unstable shapes, respectively.

The IBM-1 model describes a system of the $s$- and the $d$-bosons interactions with angular
momenta L=0 and L=2, respectively. The model Hamiltonian can be written in the form of
combinations of the $s$, $s^{\dag}$, $d$, $d^{\dag}$ operators~\cite{Iac87} of the $s$-
and $d$-bosons. The Hamiltonian is also written in terms of Casimir operators of the
dynamical symmetries. For the present application, the multipole form of Hamiltonian~\cite{Casten88}
was selected as
\begin{equation}
	\hat H=
	\epsilon\,\hat n_d+
	\kappa\,\hat Q\cdot\hat Q+
	\kappa'\,\hat L\cdot\hat L,
	\label{e_ham}
\end{equation}
where $\hat n_d$, $\hat Q$, $\hat L$ are the boson-number, quadrupole and
angular momentum operators, respectively,  defined  in the form of
combination of the operators $s$, $s^{\dag}$, $d$, $d^{\dag}$
\begin{eqnarray}
	\hat n_d&=&\sqrt{5}[d^\dag\times\tilde d]^{(0)}_0,
	\nonumber\\
	\hat Q&=&[d^\dag\times\tilde s+s^\dag\times\tilde d]^{(2)}+
	\chi[d^\dag\times\tilde d]^{(2)},
	\nonumber\\
	\hat L&=&\sqrt{10}[d^\dag\times\tilde d]^{(1)}.
	%\nonumber\\
	\label{e_term}
\end{eqnarray}
The constants $\epsilon$, $\kappa$, $\kappa'$ in Eq.~(\ref{e_ham}), and $\chi$ associated with the quadrupole operator $\hat Q$, serve as free parameters in the Hamiltonian. These parameters are adjustable and can be tuned to match the experimental data~\cite{NNDC23} to calculate the
energy levels of a given nucleus. Besides energy level calculation,
the geometric shape of nuclei can also be predicted
by plotting the energy surface as a function of the deformation
parameters. The PES as a function of $\beta$ and $\gamma$, obtained from the model
Hamiltonian~(\ref{e_ham}) in the classical limit~\cite{Dieperink80,Ginocchio80}, 
can be formulated as follows
{\scriptsize
	\begin{eqnarray}%\begin{equation}%\begin{split}
		V(\beta,\gamma)=
		\epsilon_d \frac {N \beta^2}{1+\beta^2} + \kappa' 6N\frac{\beta^2}{1+\beta^2}
		+ \kappa \frac{ N }{1+\beta^2} \nonumber\\
		\times \left[ 5 + (1 + \chi^2) \beta^2
		+ \frac{(N-1)\left(\frac{2\chi^2\beta^2}{7}-4\sqrt{\frac{2}{7}}\chi\beta\cos3\gamma+4\right)\beta^2}{1+\beta^2}\right],
		\label{pes}%\end{split}
	\end{eqnarray}%\end{equation}
}where $N$ is number of the bosons. The $\epsilon$, $\kappa$, $\kappa'$ and $\chi$ are common parameters given as constants in Eqs.~(\ref{e_ham}) - (\ref{e_term}).
The $\beta$ and $\gamma$ variables are called deformation parameters and are the same as introduced in the collective model of Bohr and Mottelson~\cite{Boh98}.
Both parameters ($\beta$ and $\gamma$) are $zero$ for spherical nuclei while $\beta\neq0$, $\gamma=0^\circ$, and
$60^\circ$ for prolate and oblate nuclei, respectively.

\subsection{Relativistic Mean Field (RMF) Model}
%%%%%%%%%%%%%%%%%%%%%
In RMF model,  a nucleus is described as a system of Dirac nucleons coupled to exchange of various mesons such as isoscalar scalar $\sigma$ meson, the isoscalar vector $\omega$ meson, and the isovector vector  $\rho$ meson together with the electromagnetic field~\cite{serot86,serot97,Boguta1977}. In addition, a treatment of nuclear matter and finite nuclei requires a medium dependence of effective mean-field interactions. This dependence can either be introduced by assuming an explicit density dependence for the meson-nucleon couplings or by including non-linear meson self-interaction terms in the phenomenological Lagrangian density \cite{nik}. One can find many approaches for the construction of successful phenomenological non-linear RMF interactions (i.e NL3, PK1, PK1R FSUGold).  Later density-dependent versions of RMF model were introduced. The density dependence of the meson-nucleon vertex functions can be parameterized either from microscopic calculations of nuclear matter or adjusting data and empirical properties of finite nuclei. Meson exchange is the convenient effective nuclear interaction for the ground state and low-lying  excited states of finite nuclei. However, exchange of heavy mesons is associated with the short distance dynamics which means that it cannot be resolved at low energies. Because of this reason,  density-dependent point coupling version of RMF model was introduced. In this version, meson exchanges are replaced by the corresponding local four-point interactions between nucleons. More details and discussions of the versions of the RMF model can be found in Refs.~\cite{Lalazissis97,Ring1996,Typel1999,vretenar2005,Meng2006,Haddad2007,maza11,typel18}.

In the current investigation, we have used effective density-dependent meson-exchange functional, DD-ME2~\cite{lalazissis2005} and density-dependent point-coupling functional, DD-PC1~\cite{niksic2008} for RMF calculations. It should be noted that we have restricted ourselves to describing only the meson-exchange version of the RMF model in this subsection.

In the density-dependent meson-exchange framework of RMF model, the nucleus is defined as a set of Dirac nucleons interacting through the exchange of mesons, resulting in finite-range interactions~\cite{Typel1999}. The Lagrangian equation of the meson-exchange version of the model can be divided into three parts as follows

\begin{equation}    \mathcal{L}=\mathcal{L}_N+\mathcal{L}_m+\mathcal{L}_{int}\label{lagpart},
\end{equation}
where $\mathcal{L}_N$, $\mathcal{L}_m$ and $\mathcal{L}_{int}$ terms are related with free nucleon, fields of the free meson and electromagnetic fields, and meson-nucleon interactions, respectively~\cite{Ring1996,Meng2006}.
%The dynamics of the nucleon interacting with each other through the exchange of mesons are represented by the Lagrangian of free nucleon
Open form of $\mathcal{L}_N$ term is given by
\begin{equation}
	\mathcal{L}_N=\bar{\Psi}(i\gamma_\mu\partial^\mu-m)\Psi, \label{Lagrangian2}
\end{equation}
where the Dirac spinors ($ \bar{\Psi} $ and $ \Psi $), m, and $\gamma_\mu$ represent the nucleon field, meson field, nucleon mass, and Dirac gamma matrices, sequentially. Lagrangian density for the meson and electromagnetic fields ($\mathcal{L}_m$) is given as
\begin{equation}
	\begin{split}
		\mathcal{L}_m &=\frac{1}{2}\partial_\mu\sigma\partial^\mu\sigma-\frac{1}{2}m_\sigma^2\sigma^2-\frac{1}{2}\Omega_{\mu\nu}\Omega^{\mu\nu}+\frac{1}{2}m^2_\omega\omega_\mu\omega^\mu \\
		&-\frac{1}{4}\overrightarrow{R}_{\mu\nu}.\overrightarrow{R}^{\mu\nu}+\frac{1}{2}m_\rho^2 \overrightarrow{\rho}_\mu.\overrightarrow{\rho}^\mu -\frac{1}{4}F_{\mu\nu}F^{\mu\nu}. \label{lagrangian3}
	\end{split}
\end{equation}
This term includes the kinetic energy of the meson field, potential energies of the $\sigma$, $\omega$ and $\rho$ mesons and the electromagnetic field. In Eq.~(\ref{lagrangian3}), arrows represent isovectors and $m_\sigma$, $m_\omega$ and $m_\rho$ are the mass of $\sigma$, $\omega$ and $\rho$ mesons, respectively. Field tensors are given by the following equations
\begin{equation}
	\begin{split}
		\Omega_{\mu\nu}=\partial_\mu\omega_\nu-\partial_\nu\omega_\mu, \\
		\overrightarrow{R}_{\mu\nu}=\partial^\mu\overrightarrow{\rho}_\nu-\partial_\nu\overrightarrow{\rho}_\mu, \\
		F_{\mu\nu}=\partial_\mu A_\nu-\partial_\nu A_\mu  \label{fieldtensors}.
	\end{split}
\end{equation}
In Eq.~(\ref{lagpart}),  $\mathcal{L}_{int}$  is the interaction term, which refers to the interaction between mesons and nucleons, given by
\begin{equation}
	\begin{split}
		\mathcal{L}_{int}= -g_\sigma\bar{\Psi}\Psi \sigma - g_\omega\bar{\Psi}\gamma^\mu\Psi\omega_\mu -&\\ g_\rho\bar{\Psi}\overrightarrow{\tau}\gamma^\mu\Psi.\overrightarrow{\rho}_\mu-e\bar{\Psi}\gamma^\mu\Psi A_\mu,
	\end{split}
	\label{Lagrangian4}
\end{equation}
where the coupling constants of the $\sigma$, $\omega$ and $\rho$ mesons are defined by $g_\sigma$, $g_\omega$ and $g_\rho$, sequentially~\cite{Ring1996,vretenar2005,Meng2006,Haddad2007}

A  Hamiltonian density of the static case can be calculated from the Lagrangian density. The total energy ($E_{RMF}$), which depends on the Dirac spinors and the meson fields, is obtained by integrating Eq.~(\ref{lagpart}) over the r-space.

\begin{equation}
	E_{RMF}\left[\Psi,\bar{\Psi},\sigma,\omega^{\mu},\overrightarrow{\rho}^{\mu}, A^{\mu} \right]=\int d^3r \mathcal{H}({\bf r}). \label{toten}
\end{equation}

The density-dependent meson-exchange model incorporates an explicit density dependence for the meson-nucleon vertices. To determine the properties of finite nuclei, the meson-nucleon vertex functions can be adjusted by tuning the parameters of the meson-nucleon couplings' density dependence~\cite{lalazissis2005} . 
%The couplings of $\sigma$ and $\omega$ mesons to the nucleon field are expressed as
%	\begin{equation}
	%		g_i(\rho) = g_i(\rho_{sat})f_i(x) ~~~\text{for}~i=\sigma, \omega. \label{coup1}
	%	\end{equation}
%In Eq.~(\ref{coup1}), $\rho_{sat}$ represents the baryon density at saturation in symmetric nuclear matter whereas the function $f_i(x)$ is defined as
%	\begin{equation}
	%		f_i(x) = a_i\frac{1 + b_i(x + d_i)^2}{1 + c_i(x + d_i)^2}, \label{coups}
	%	\end{equation}
%where $x = \rho/\rho_{sat}$, and the eight parameters in Eq.~(\ref{coups}) are subject to constraints: $f_i(1) = 1$, $f_\sigma(1)^{''} = f_\omega^{''}(1)$, and $f_i^{''}(0) = 0$. These constraints reduce the number of parameters for density dependence to three. The density dependence of the $\rho$ meson is described by an exponential form
%	\begin{equation}
	%		g_\rho(\rho) = g_\rho(\rho_{sat})e^{-a_\rho(x - 1)}. \label{coup}
	%	\end{equation}
%In this case, $\rho(\rho_{sat})$ and $a_\rho$ provide a parametrization of the isovector channel. To determine the properties of finite nuclei, the meson-nucleon vertex functions can be adjusted by tuning the parameters of the meson-nucleon couplings' density dependence.

In this paper, we have used effective density-dependent meson-exchange DD-ME2 and point-coupling DD-PC1 functionals for calculation of ground-state binding energies per nucleon and determination of potential energy surfaces (PESs) of the selected nuclei. On the other hand, pairing correlations is an important physical quantity in investigations of open shell nuclei. Generally, pairing has been considered in the Bardeen–Cooper–Schrieffer (BCS) approach phenomenologically by regarding monopole pairing force, adjusted to the experimental mass differences of odd–even nuclei which can be a poor approximation in many cases. Because of this reason, a new formulation for pairing has been considered in the relativistic Hartree–Bogoliubov (RHB) model where the particle–particle channel of the effective inter-nucleon interaction is described by a separable finite range in pairing force (see Ref.~\cite{niksic2014} and references therein).  In this study, we have considered a separable finite range pairing force and followed the prescription of Ref.~\cite{niksic2014} for triaxially symmetric RMF calculations of $^{68}$Se, $^{70}$Se, $^{70}$Br, $^{70}$Kr, $^{72}$Kr, $^{74}$Kr, $^{74}$Rb and $^{74}$Sr.

%In this paper, we have used effective density-dependent meson-exchange DD-ME2 and point-coupling DD-PC1 functionals for calculation of ground-state binding energies per nucleon and determination of PESs of the selected nuclei. On the other hand, pairing co-relations is an important physical quantity in investigations of open shell nuclei. Generally, pairing has been considered in the %    BardeenCooperSchrieffer (BCS) approach phenomenologically by regarding monopole pairing force adjusted to the experimental mass differences of oddeven nuclei which can be a poor approximation in many cases. Because of this reason, a new formulation for pairing has been considered in the relativistic HartreeBogoliubov (RHB) model where the particleparticle channel of the effective inter-nucleon interaction is described by a separable finite range in pairing force (see Ref.~\cite{niksic2014} and references therein).  In this study, we have considered a separable finite range pairing force and followed the prescription of Ref.~\cite{niksic2014} for triaxially symmetric RMF calculations of $^{68}$Se, $^{70}$Se, $^{70}$Br, $^{70}$Kr, $^{72}$Kr, $^{74}$Kr, $^{74}$Rb and $^{74}$Sr.}
%...................
%%%%%%%%%%%%%%%%%%%%%
\subsection{The proton-neutron Quasiparticle Random Phase Approximation (pn-QRPA) Model}
The $\beta$-decay properties were studied within the quasiparticle random phase approximation with a separable multi-shell interaction on top of axially symmetric deformed mean-field calculations employing Nilsson potential. The following Hamiltonian was chosen for solution in the pn-QRPA model
\begin{equation} \label{H}
H^{pnQRPA} = H^{sp} + V^{ph}_{GT} + V^{pp}_{GT} + V^{pair},
\end{equation}
where, $H^{sp}$ corresponds to the single-particle Hamiltonian, $V_{GT}^{pp}$ and $V_{GT}^{ph}$ represent the particle-particle and particle-hole GT forces, respectively. The particle-particle ($pp$) force was initially neglected for the computation of $\beta^{-}$ decay~\cite{staudt1990}. Later investigations revealed that this force is essential for an accurate determination of $\beta^{+}$ decay \cite{hirsch1993}.  The \textit{pp} and \textit{ph} interaction strengths, characterized by $\kappa$ and $\chi$, respectively were parametrized following a 1/A$^{0.7}$ dependence, as proposed in Ref.~\cite{Hom96}. The last term in Eq.~(\ref{H}) denotes the pairing force and was calculated using the BCS approach. For computing the single-particle energies and wavefunctions, the Nilsson model \cite{Nil55} with incorporation of quadrupole deformation parameter  ($\beta_2$)
was employed. The oscillator constant for nucleons was determined using $\hbar\omega=\left(45A^{-1/3}-25A^{-2/3}\right)$ MeV. The same value of oscillator constant was applied for protons and neutrons. The Nilsson-potential parameters were adopted from Ref.~\cite{ragnarson1984}.  $Q$-values were taken from Ref.~\cite{Aud21}. Traditional pn-QRPA calculations used $\Delta_p$ = $\Delta_n$ = 12/$\sqrt{A}$ MeV \cite{hirsch1993}. However, recent findings~\cite{Ull23} revealed that the three-term formula, based on neutron and proton separation energies, resulted in overall best prediction of $\beta$-decay half-lives using the current pn-QRPA model. Pairing gaps for proton and neutron were computed using separation energies of proton (S$_p$) and neutron (S$_n$), respectively, as follows
\begin{eqnarray}
\bigtriangleup_{pp} =\frac{2}{8}(-1)^{Z+1}[S_p(A+1, Z+1)+S_p(A-1, Z-1)-2S_p(A, Z)]
\end{eqnarray}
\begin{eqnarray}
\bigtriangleup_{nn} =\frac{2}{8}(-1)^{A-Z+1}[S_n(A+1, Z) + S_n(A-1, Z)- 2S_n(A, Z)].
\end{eqnarray}
The reduced GT transition probabilities was calculated using
\begin{equation}
B_{GT}(E_j) = \frac{1}{2J_i +1}\mid\langle j \parallel M_{GT} \parallel i \rangle\mid^2,
\end{equation}
\begin{equation}
M_{GT} = \sum_{k,\mu} \tau_+(k) \sigma_{\mu}(k), \nonumber
\end{equation}
where $\tau_+$ and $\sigma_{\mu}(k)$ are the iso-spin raising and spherical components of the spin operator, respectively. $E_j$ are the energy levels in daughter and $\mu$ = (-1, 0, 1) denote the third component of the angular momentum of the nucleons.

The calculation of $\beta$-decay partial half-lives was performed using
\begin{eqnarray}
	t_{1/2}^{p} = \frac{C}{(g_V/g_A)^{2}f_A(A, Z, E)B_{GT}(E_j)+f_V(A, Z, E)B_F(E_j)},
\end{eqnarray}
where $E$ = ($Q$ - $E_j$),  $g_A/g_V$ = -1.254 and $C$ = 6295 $s$. $f_{A/V}$ are the Fermi integral functions for axial vector and vector transitions. $B_{GT}$ ($B_{F}$) stands for the reduced transition probability for the GT (Fermi) transitions.
The total $\beta$-decay half-lives were computed using
\begin{equation}
T_{1/2} = \left(\sum_{0 \le E_j \le Q} \frac{1}{t_{1/2}^{p}}\right)^{-1}.
\end{equation}
For details on complete solution of Eq.~(\ref{H}),  not reproduced here for space consideration,  we refer to~\cite{hirsch1993}.

\section{Results and Discussions}
The current investigation explores the effect of quadrupole deformation parameters ($\beta_2$) on $\beta$-decay observables. To do the needful,  four different  deformation parameters were computed using the RMF model using the DD-ME2 and DD-PC1 functionals. Each functional yielded two local minima in the PESs, one each for oblate (O) and prolate (P) shapes. These are referred to as $\beta_2$(DD-ME2 (O)), $\beta_2$(DD-ME2 (P)), $\beta_2$(DD-PC1 (O)) and $\beta_2$(DD-PC1 (P)), respectively. The IBM-1 model yielded spherical shapes for all selected nuclei and are denoted by $\beta_2$(IBM-1 (S)).  In addition, two more  $\beta_2$ were included in our investigation. The FRDM~\cite{Mol16} computed deformation values are denoted by $\beta_2$(FRDM) and measured deformation are denoted by $\beta_2$(NNDC) taken from Ref.~\cite{NNDC23}. 

\begin{figure}[h]
	\centering
	\includegraphics[width=1\textwidth]{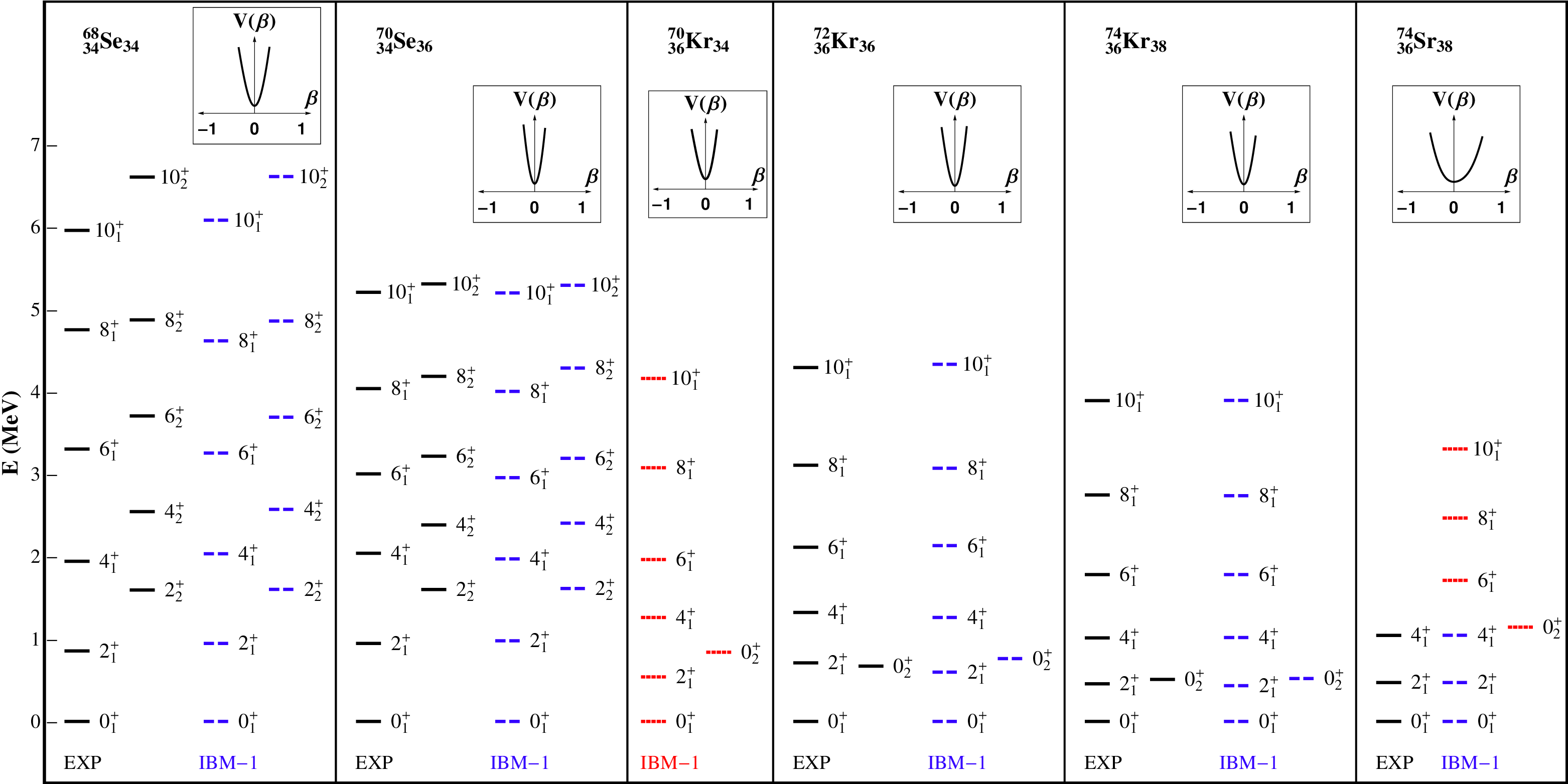}
	\caption{The experimental (solid), calculated (dashed) and predicted (dotted) energy spectra of $^{68}$Se, $^{70}$Se, $^{70}$Kr, $^{72}$Kr, $^{74}$Kr, $^{74}$Sr. The inset shows PESs as a function of $\beta$ for $\gamma$ = $0^\circ$.} \label{enpes}
\end{figure}

For the calculation of energy levels of even-even nuclei in the $A\sim70$ region, the multipole form of \mbox{IBM-1} Hamiltonian given
in Eq.~(\ref{e_ham}) was used. There are four free parameters $\epsilon$, $\kappa$,
$\kappa'$  and $\chi$ in the model.
Initially, Hamiltonian parameters were
fitted for $^{68}$Se and later expanded for heavier nuclei up to $^{74}$Sr. In order to perform the fitting procedure, $\epsilon$ was first arranged.  {Later, $\kappa$, $\kappa'$ and $\chi$ parameters} were determined by achieving a best-fit with the experimental data~\cite{NNDC23}. The set of fitted parameters are given in Table~\ref{par}.  The low-lying
energy spectra of selected nuclei are exhibited in Fig.~\ref{enpes}. It may be noted from the figure that the calculated
results are in good agreement with the experimental data~\cite{NNDC23}. The energy
levels of unknown $^{70}$Kr isotope were predicted by using the set of parameters of $^{72}$Kr. {Similarly, unknown} energy levels in the ground state band of $^{74}$Sr were predicted and shown as dotted lines in Fig.~\ref{enpes}. The PESs as a function of deformation parameters ($\beta$, $\gamma$) for each nucleus are depicted in top panels of Fig.~\ref{enpes}. These PESs were plotted by using the common set of fitted parameters given in Table~\ref{par}. It is noted that IBM-1 predicts spherical shapes for all nuclei. Accordingly  their deformation
{parameters %($\beta_2$) 
are} determined as $zero$.

The non-linear version of the RMF model was used to calculate various fundamental ground-state
properties of nuclei including binding energy per nucleon (BE/A), nucleon separation energies, nuclear
charge radii, deformations and electric moments throughout the nuclear chart~\cite{bayram13a,lalazissis1999}.
The RMF model, with a small number of adjustable parameters,  provides the correct
prediction of ground-state energies, sizes, and deformations of nuclei~\cite{Ring1996}.
In the current investigation, we present the RMF model calculation, with density-dependent
forces, for $BE/A$, two-nucleon separation energies and $\beta_2$ values of selected neutron-deficient nuclei.

\begin{figure}[h]
	\centering
	\includegraphics[width=0.9\textwidth]{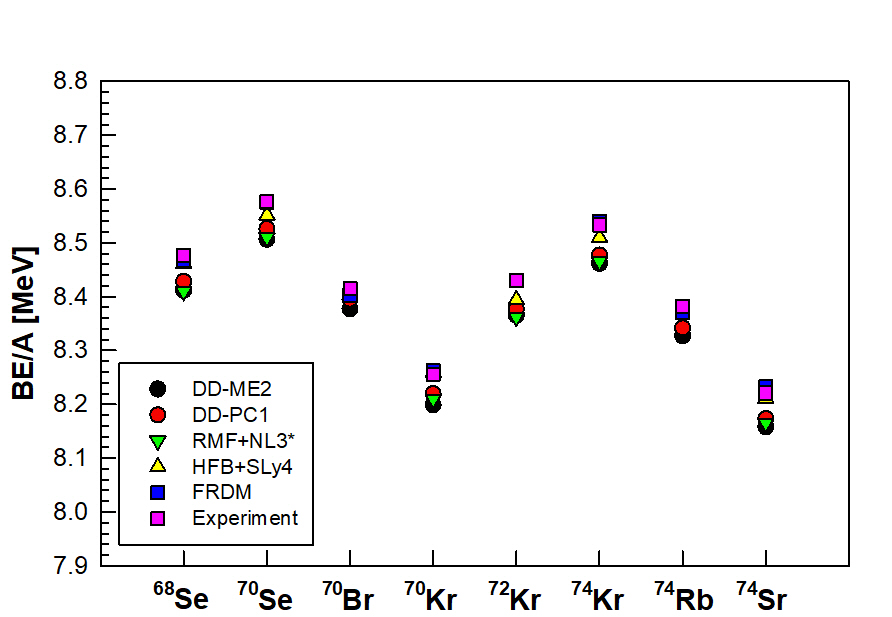}
	\caption{The calculated BE/A values with DD-ME2 and DD-PC1 functionals in comparison with the results of RMF model with NL3* interaction~\cite{bayram13a}, HFB method with SLy4 force~\cite{stoitsov2003}, FRDM~\cite{Mol16} and experimental data~\cite{wang2017}.} \label{bea}
\end{figure}

In Fig.~\ref{bea}, the calculated BE/A values of the selected nuclei with DD-ME2 and DD-PC1 functionals are presented. Our results are compared with RMF model with NL3* interaction~\cite{bayram13a}, Hartree-Fock Bogoliubov (HFB) method with SLy4 parameter set~\cite{stoitsov2003}, Finite Range Droplet Model (FRDM)~\cite{Mol16} and experimental data~\cite{wang2017}. All theoretical models predict $BE/A$ values consistent with experimental data. The best theoretical results were obtained with the FRDM data. The maximum deviation of FRDM from experimental data is 0.006 MeV while those of other models go up to 0.06 MeV.  Root mean square errors (RMSE) between model prediction values and experimental data are (0.061, 0.045, 0.062, 0.022, 0.008) MeV for RMF+DD-ME2, RMF+DD-PC1, RMF+NL3*, HFN+Sly4 and FRDM, respectively. It should be noted that HFB and RMF model results were obtained with a smaller number of adjustable parameters.

%The percentage uncertainty between the predicted values and experimental data is 0.08, 0.06, 0.08, 0.03, and 0.009 for RMF+DD-ME2, RMF+DD-PC1, RMF+NL3*, HFN+Sly4 and FRDM, respectively.. It should be noted that HFB and RMF model results were obtained with a smaller number of adjustable parameters.

\begin{figure}[h]
	\centering
	\includegraphics[width=120mm]{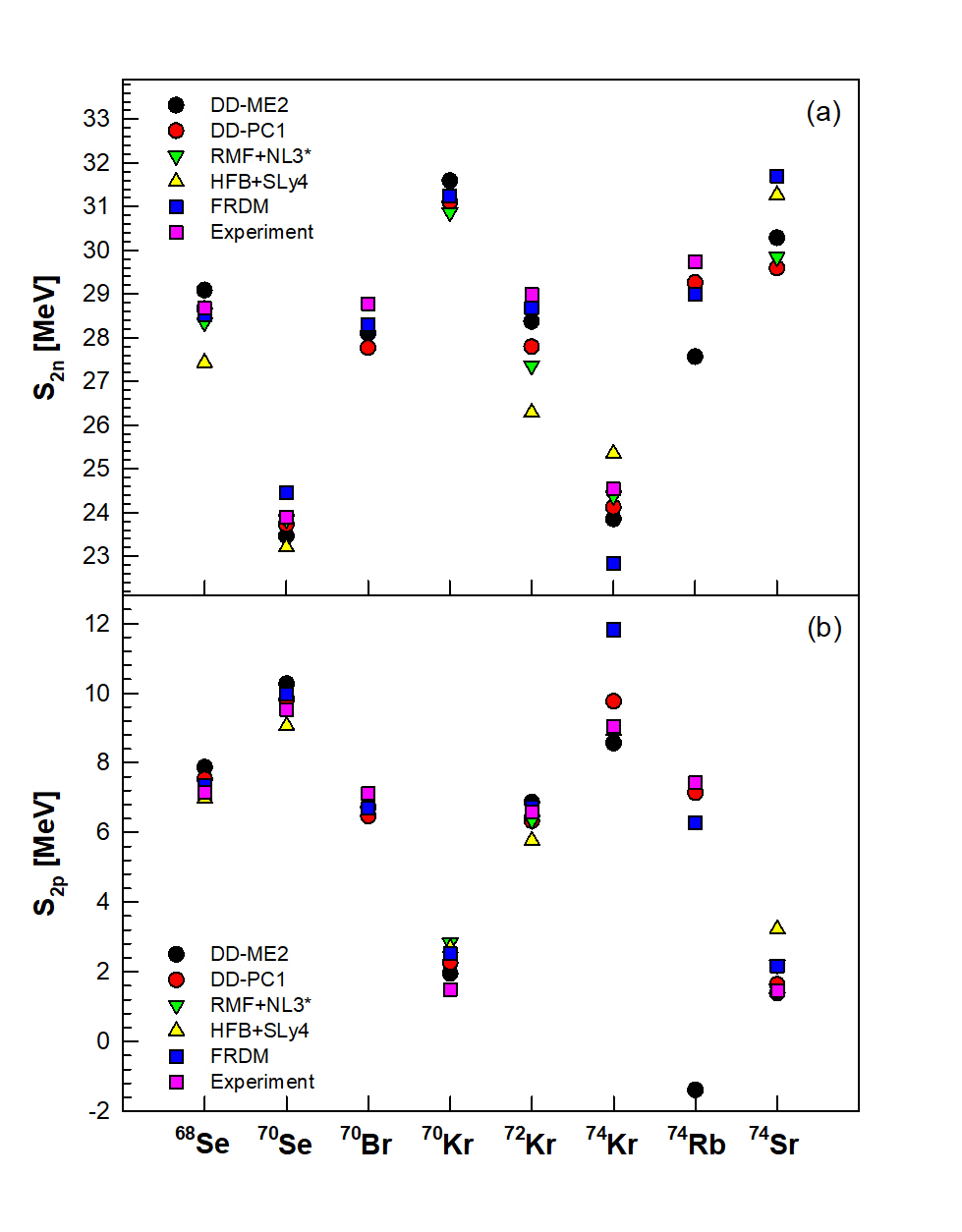}
	\caption{The calculated two-neutron (a) and two-proton (b) separation energies of selected nuclei. Shown also are the results of RMF model with NL3* interaction~\cite{bayram13a}, HFB method with SLy4 force~\cite{stoitsov2003}, FRDM~\cite{Mol16} and experimental data~\cite{wang2017}, wherever applicable.} \label{sep}
\end{figure}

We computed two-neutron (S$_{2n}$) and two-proton (S$_{2p}$) separation energies of selected nuclei  using the calculated $BE$ values. In our calculations, we  used $S_{2n}=BE(Z,N)-BE(Z,N-2)$ and $S_{2p}=BE(Z,N)-BE(Z-2,N)$ for determination of two-nucleon separation energies.  Fig.~\ref{sep} presents the calculated $S_2n$ (a) and $S_{2p}$ (b) values of $^{68}$Se, $^{70}$Se, $^{70}$Br, $^{70}$Kr, $^{72}$Kr, $^{74}$Kr, $^{74}$Rb and $^{74}$Sr  together with available theoretical and experimental data. It is noted from Fig.~\ref{sep}(a) and~\ref{sep}(b), all theoretical estimates for $S_{2n}$ and $S_{2p}$ are in good agreement. The HFB method with SLy4 force~\cite{stoitsov2003} predicted values show a small deviation from the measured data.

\begin{figure}[h]
	\centering
	\includegraphics[width=1\textwidth]{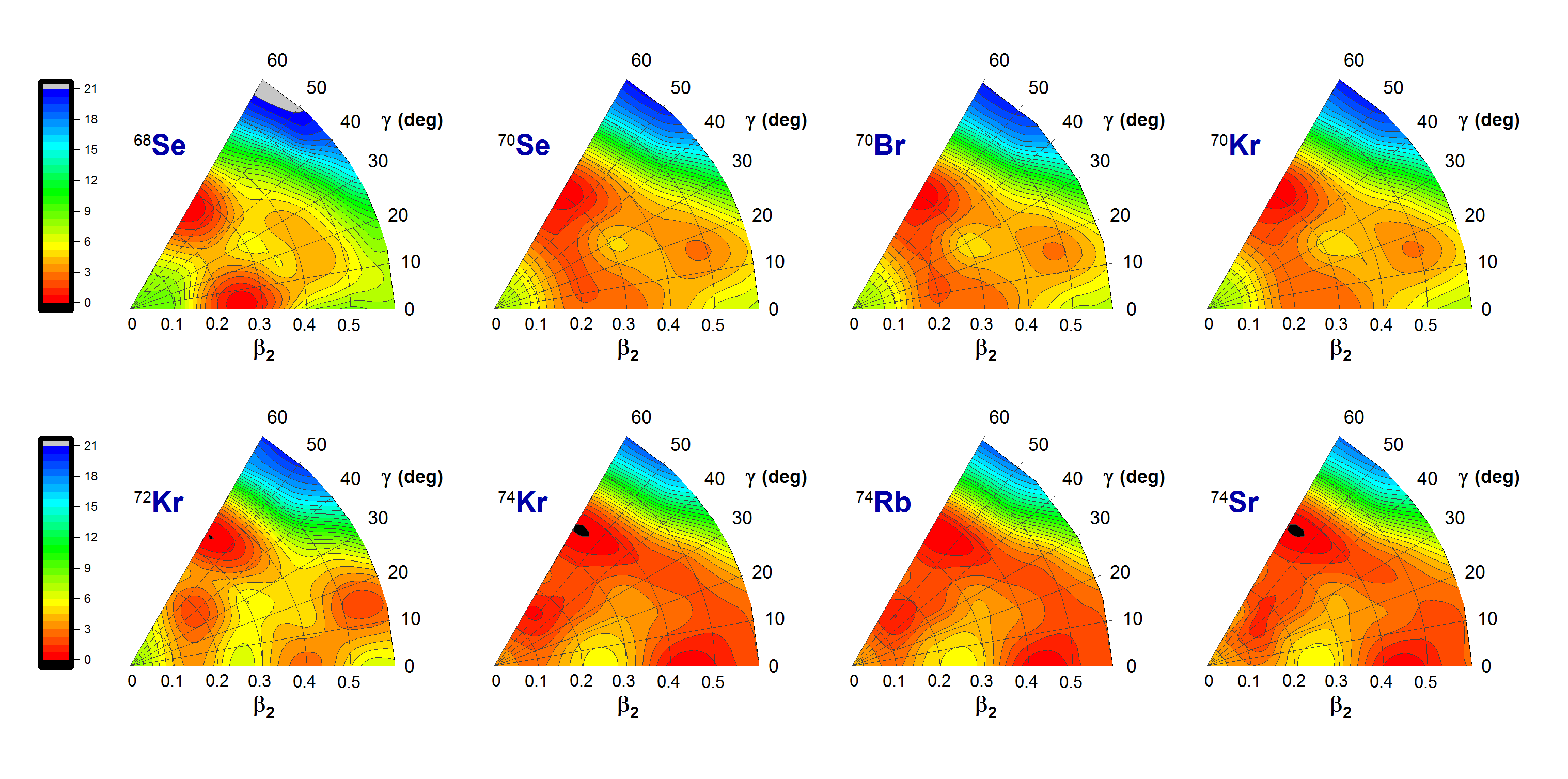}
	\caption{The PESs of $^{68}$Se, $^{70}$Se, $^{70}$Br, $^{70}$Kr, $^{72}$Kr, $^{74}$Kr, $^{74}$Rb and $^{74}$Sr  obtained from DD-ME2 calculation. See text for further details.} \label{me2}
\end{figure}

\begin{figure}[h]
	\centering
	\includegraphics[width=1\textwidth]{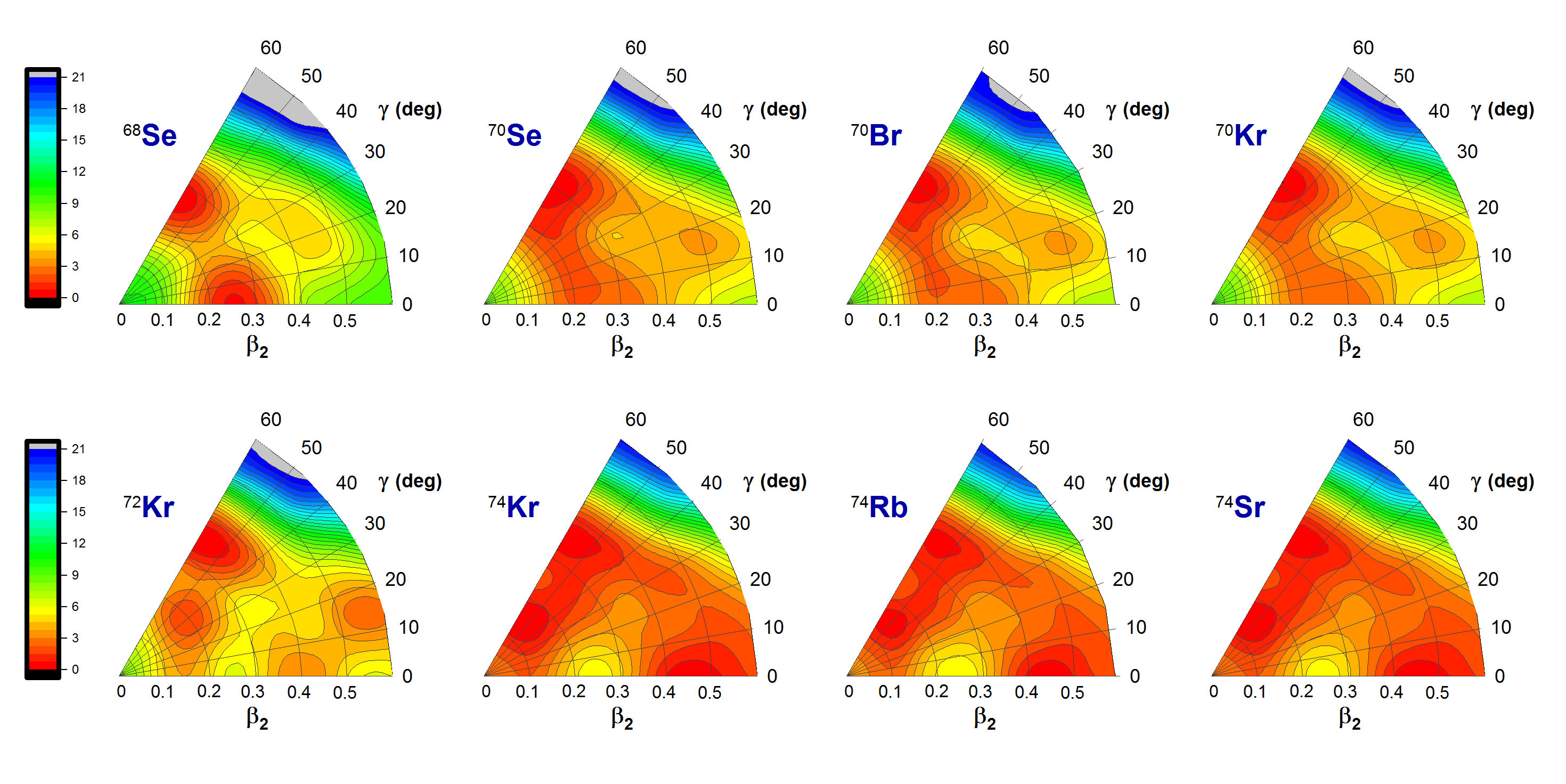}
	\caption{Same as Fig.~\ref{me2} but for DD-PC1 interaction.} \label{pc1}
\end{figure}

One of the important properties of nuclei is their ground-state deformation. We employed axially deformed self-consistent RMF calculations with DD-ME2 and DD-PC1 density-dependent interactions to investigate oblate and prolate shape configurations. The PESs quantitatively determine the ground-state shape of the nuclei. The PESs of $^{68}$Se, $^{70}$Se, $^{70}$Br, $^{70}$Kr, $^{72}$Kr, $^{74}$Kr, $^{74}$Rb and $^{74}$Sr isotopes were obtained by applying the constrained triaxially symmetric RMF model calculations. The computed PESs with DD-ME2 and DD-PC1 functionals are shown in Fig.~\ref{me2} and Fig.~\ref{pc1}, respectively. In both figures, the PESs of selected nuclei are presented on a $\beta_2-\gamma$ plane ($0 \leq \gamma \leq  60^o$). The binding energy was set to zero at the minimum of each surface, whereas the peripheral lines represent a step of 0.75 MeV. The PESs determined using the DD-ME2 and DD-PC1 interactions are  similar to each other. In PESs of $^{70}$Se, $^{70}$Br, $^{70}$Kr and $^{72}$Kr, the minimum energy configuration was obtained on the $\gamma$-axis. This translates to oblate shape prediction for these nuclei.  On the other hand, two minimum energy configurations, one on the $\gamma$-axis and the one on $\beta_2$-axis,  were obtained in the PESs of $^{68}$Se, $^{74}$Kr, $^{74}$Rb and $^{74}$Sr. In these nuclei, both DD-ME2 and DD-PC1 functional produce almost equal ground-state binding energies for oblate and prolate shapes. This means that the RMF model predicts two possible shape configurations (oblate and prolate) for $^{68}$Se, $^{74}$Kr, $^{74}$Rb and $^{74}$Sr. These deformation values were later used to investigate $\beta$-decay observables which we discuss next.

\begin{figure*}[h]
	\centering
	\includegraphics[width=1\textwidth]{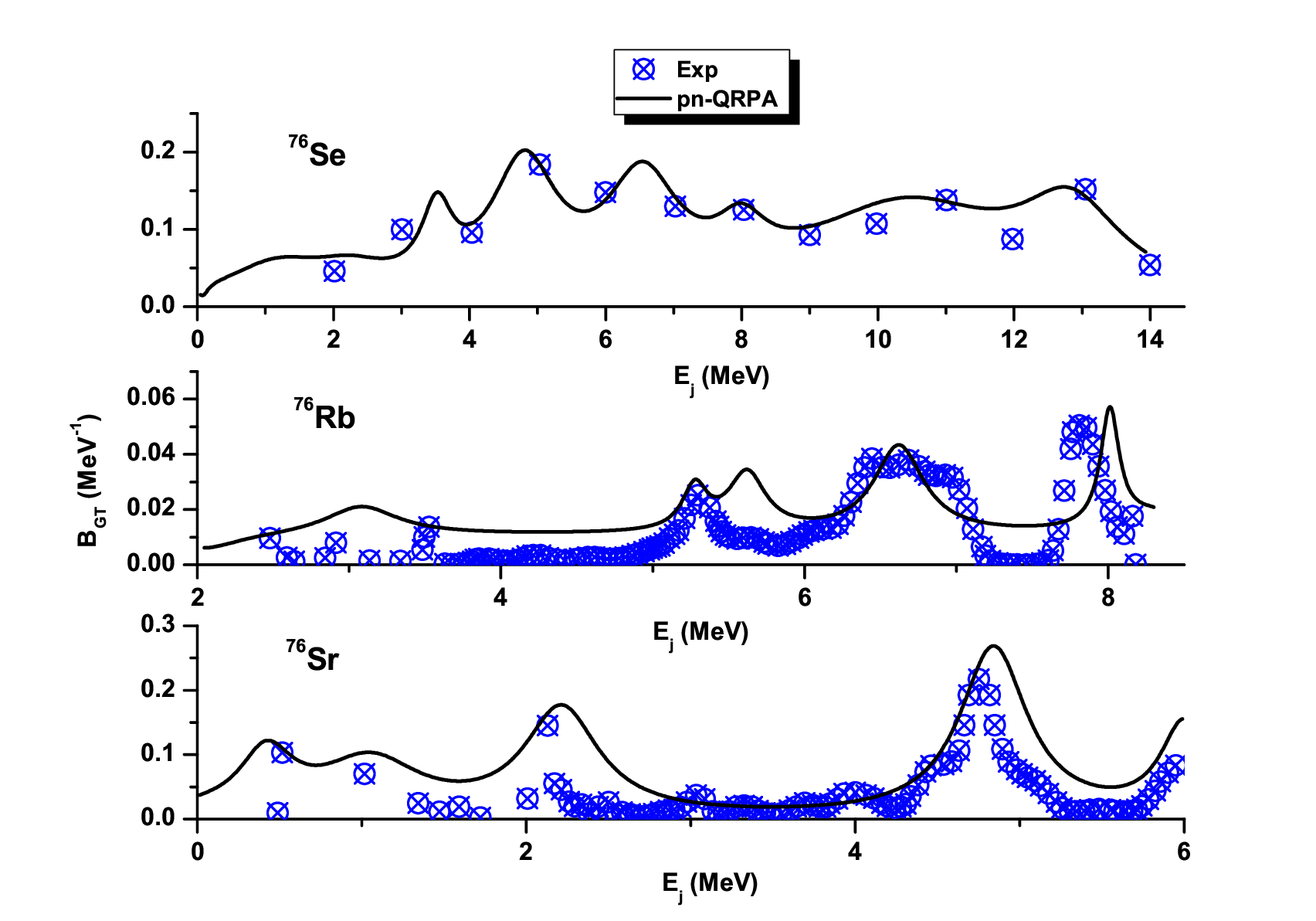}
	%	\vspace{-24mm}
	\caption{\centering Comparison of pn-QRPA calculated 
		GT$_{+}$ strength distributions of $^{76}$Se, $^{76}$Rb and $^{76}$Sr with measured data~\cite{Helmer97,Perez2013}. The abscissa shows excitation energies in daughter nuclei.} \label{g1}
\end{figure*}

\begin{figure*}[h]
	\centering
	\includegraphics[width=1\textwidth]{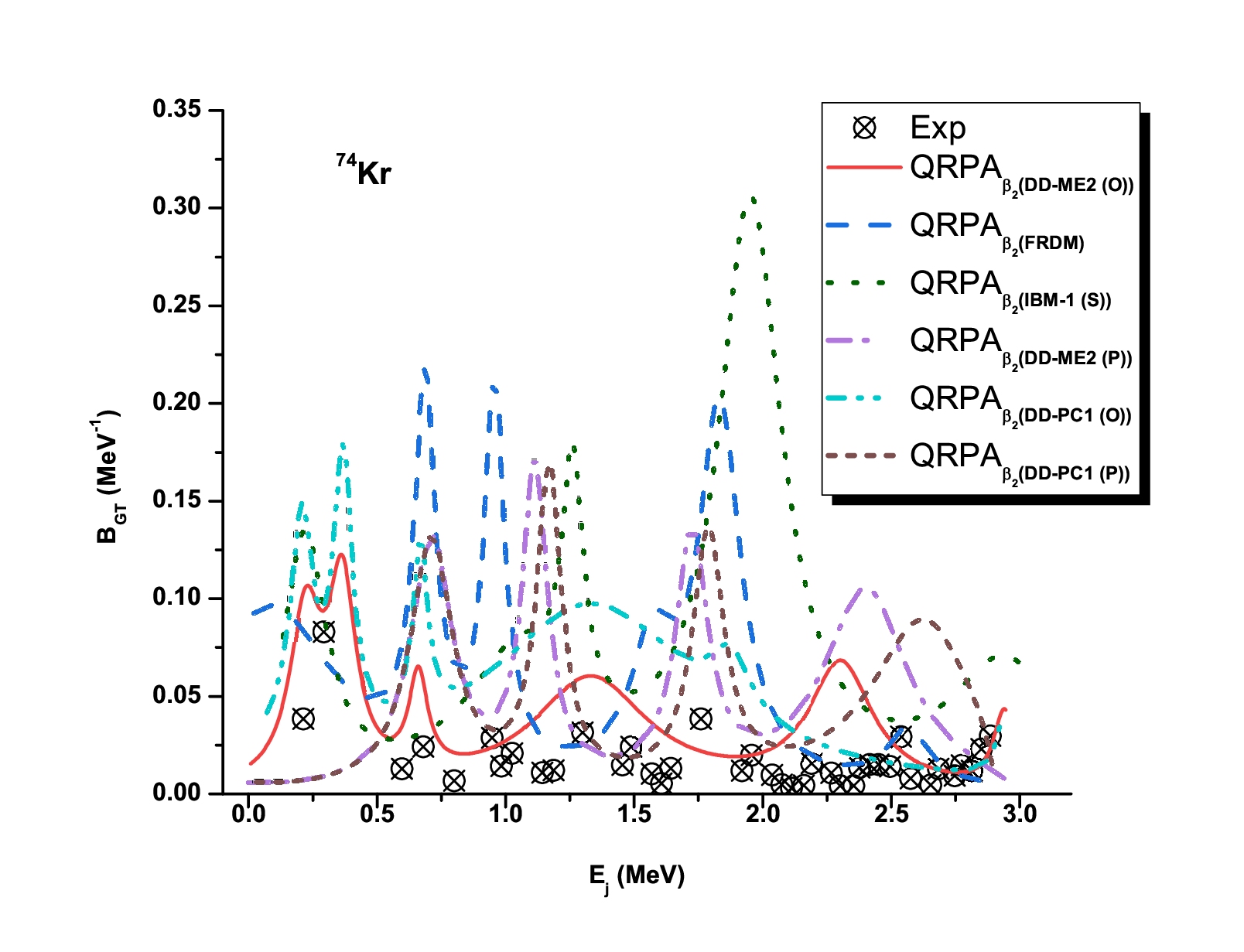}
	%	\vspace{-24mm}
	\caption{\centering Comparison of pn-QRPA calculated 
		GT$_{+}$ strength distributions of $^{74}$Kr with measured data~\cite{Poirier2004}.  The abscissa shows excitation energies in daughter nuclei. See text for explanation of symbols. } \label{g2}
\end{figure*}

The $\beta$-decaying properties of the selected nuclei were calculated in a microscopic fashion using the pn-QRPA model. We start the proceedings by presenting a comparison of our pn-QRPA calculated GT strength distributions with the measured data in order to check reliability of the current model. For most of the nuclei under study, experimental information on GT strength distributions are not available in the literature. Therefore, we were compelled to show the comparison for three other nuclei but having A$\sim$70. 
We  applied a smearing technique involving Lorentzian fitting to the theoretical strength distributions. The artificial width, applied in the fitting process, was determined based on the calculated spectrum. This widely employed technique (e.g.,~\cite{Yas18, Gue11, Wak12, Gao20}) entails comparing experimental data, measured in MeV$^{-1}$ units, with theoretically derived strength distributions. Fig.~\ref{g1} shows the comparison of measured and calculated GT distributions for $^{76}$Sr, $^{76}$Rb and $^{76}$Se. 
The measured GT distribution for $^{76}$Se was taken from Ref.~\cite{Helmer97} whereas data for $^{76}$Rb and $^{76}$Sr were taken from Ref.~\cite{Perez2013}. A decent comparison between calculated and measured GT strength distributions for $^{76}$Sr, $^{76}$Rb and $^{76}$Se  can be seen from Fig.~\ref{g1}  in (GT)$_+$ directions. 
After validation of our nuclear model, we  present our GT calculations of $^{74}$Kr  (Fig.~\ref{g2}) as a function of deformation parameter. Once again we  applied the smearing technique involving Lorentzian fitting to the theoretical strength distributions. The different calculated GT distributions are compared with the measured data~\cite{Poirier2004}. 
We denote the pn-QRPA calculated results using $\beta_2$(DD-ME2 (O)), $\beta_2$(DD-ME2 (P)), $\beta_2$(DD-PC1 (O)), $\beta_2$(DD-PC1 (P)),
$\beta_2$(IBM-1 (S)) and 
$\beta_2$(FRDM) as QRPA$_{ \beta_2 (DD-ME2 (O))}$,
QRPA$_{ \beta_2 (DD-ME2 (P))}$,
QRPA$_{ \beta_2 (DD-PC1 (O))}$,\\
QRPA$_{ \beta_2 (DD-PC1 (P))}$,
QRPA$_{ \beta_2 (IBM-1 (S))}$ and \\
QRPA$_{ \beta_2 (FRDM)}$, respectively. Calculated pn-QRPA strength distributions with input  deformation parameters from the RMF model with DD-ME2 (O) and DD-PC1 (O) functionals (namely QRPA$_{ \beta_2 (DD-ME2 (P))}$ and QRPA$_{ \beta_2 (DD-PC1 (O))}$) show good comparison with the experimental data specially for the low-lying transitions between (0 -- 0.5) MeV.

\begin{figure*}[h]
	\centering
	\includegraphics[width=1\textwidth]{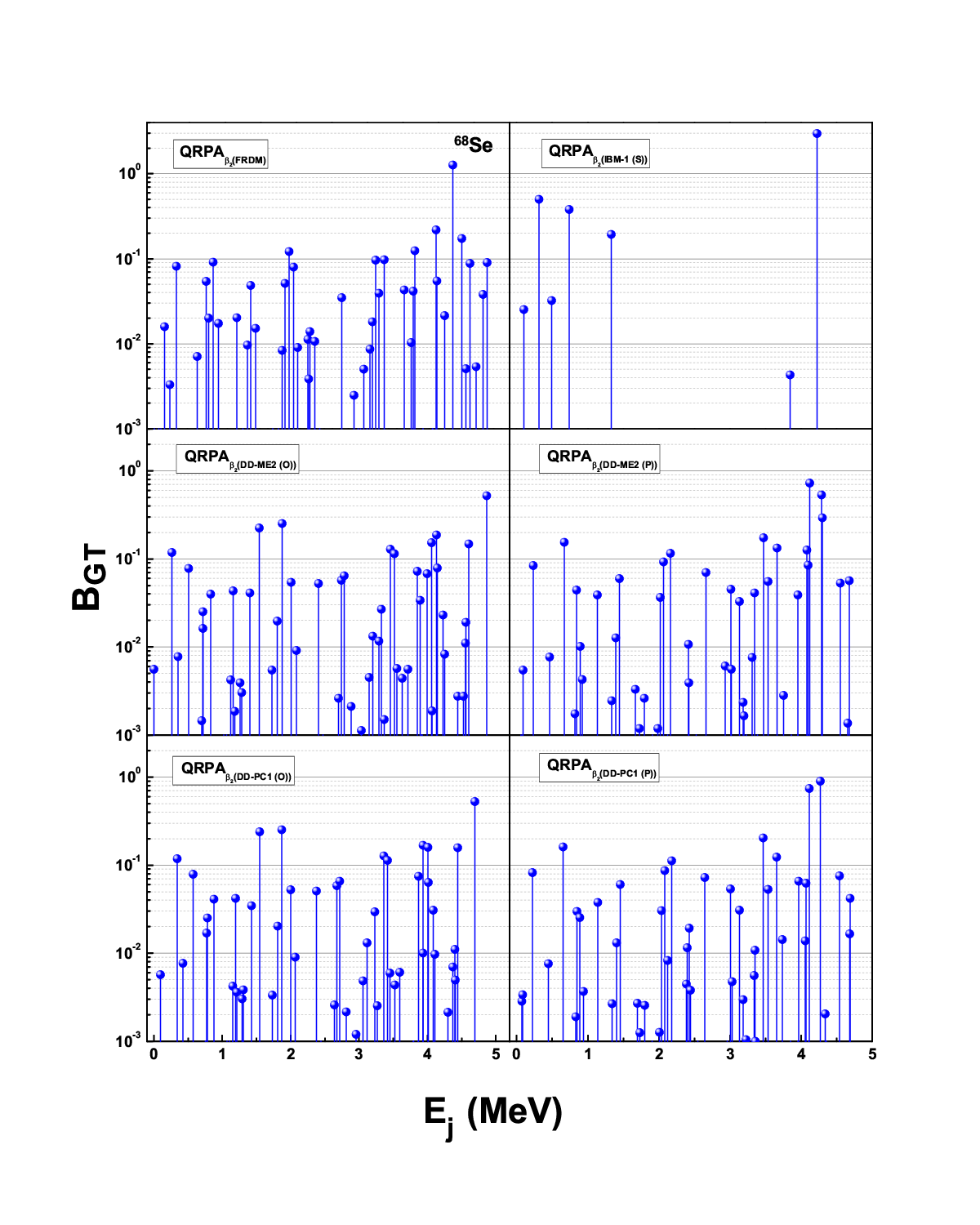}
	\vspace{-20mm}
	\caption{\centering GT strength distributions of $^{68}$Se as a function of deformation parameter obtained from models shown in the inset. The abscissa shows excitation energies in daughter nuclei.} \label{F68Se}
\end{figure*}

\begin{figure*}[h]
	\centering
	\includegraphics[width=1\textwidth]{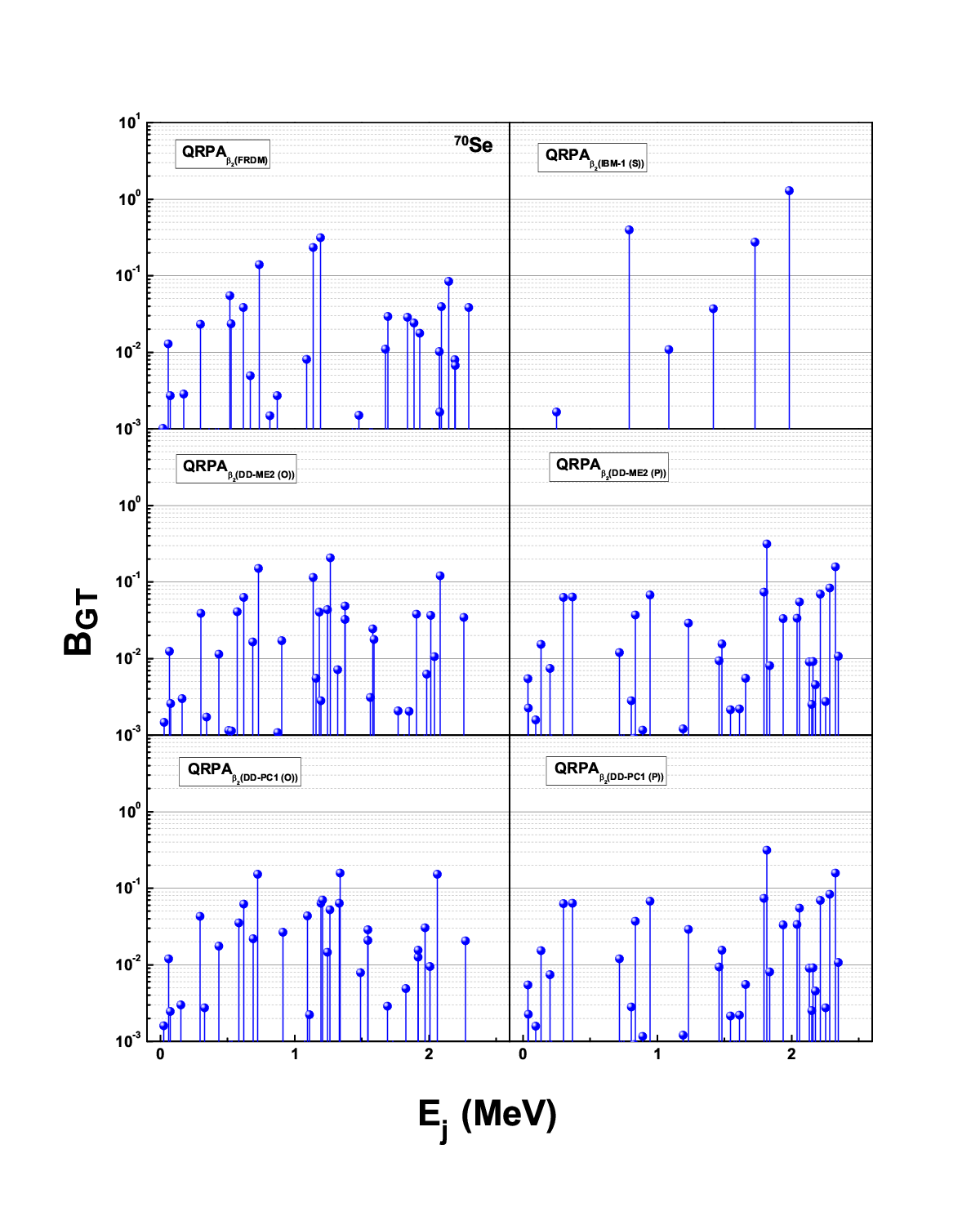}
	\vspace{-20mm}
	\caption{\centering Same as Fig.~\ref{F68Se} but for $^{70}$Se.} \label{F70Se}
\end{figure*}

\begin{figure*}[h]
	\centering
	\includegraphics[width=1\textwidth]{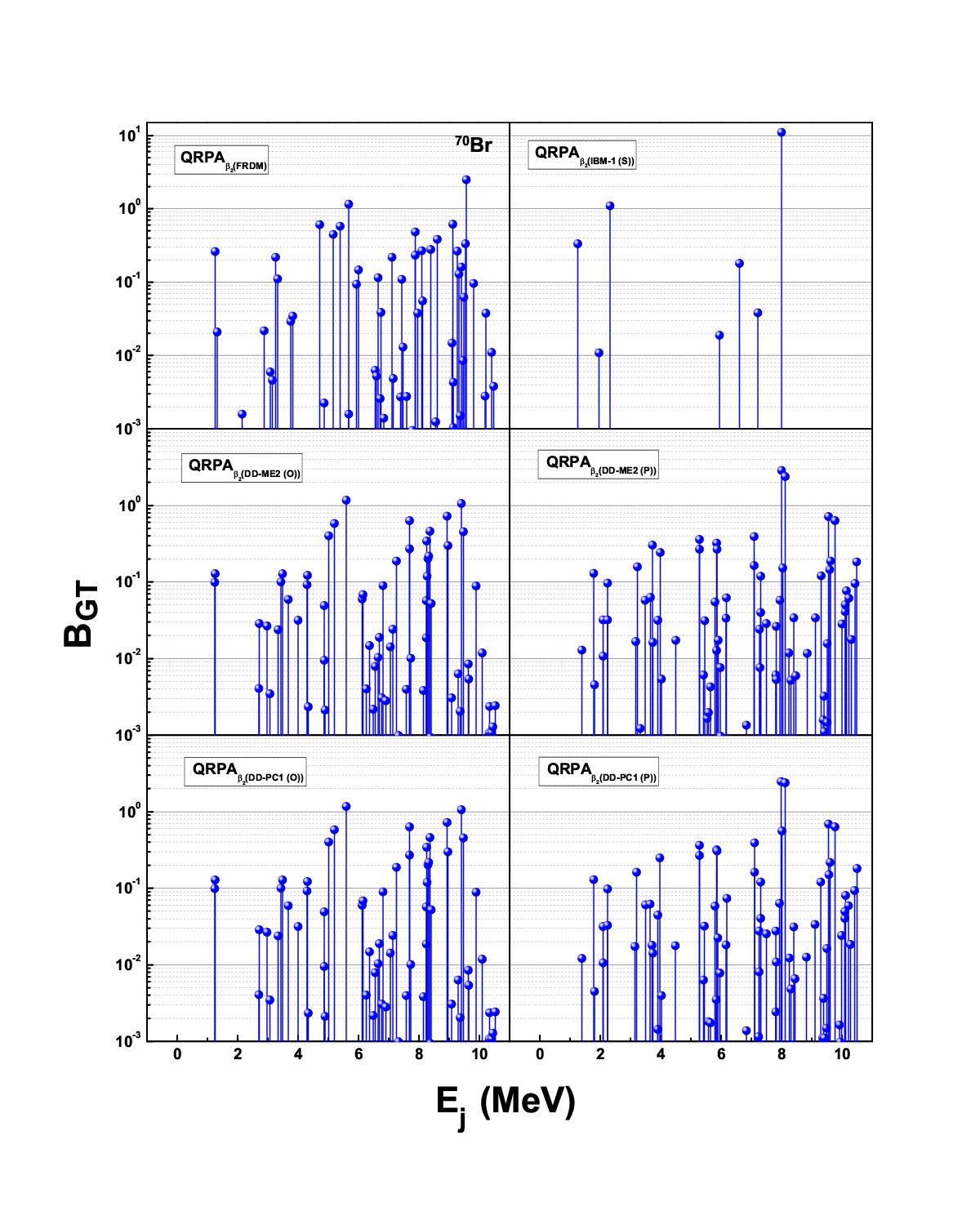}
	\vspace{-20mm}
	\caption{\centering Same as Fig.~\ref{F68Se} but for $^{70}$Br.} \label{F70Br}
\end{figure*}

\begin{figure*}[h]
	\centering
	\includegraphics[width=1\textwidth]{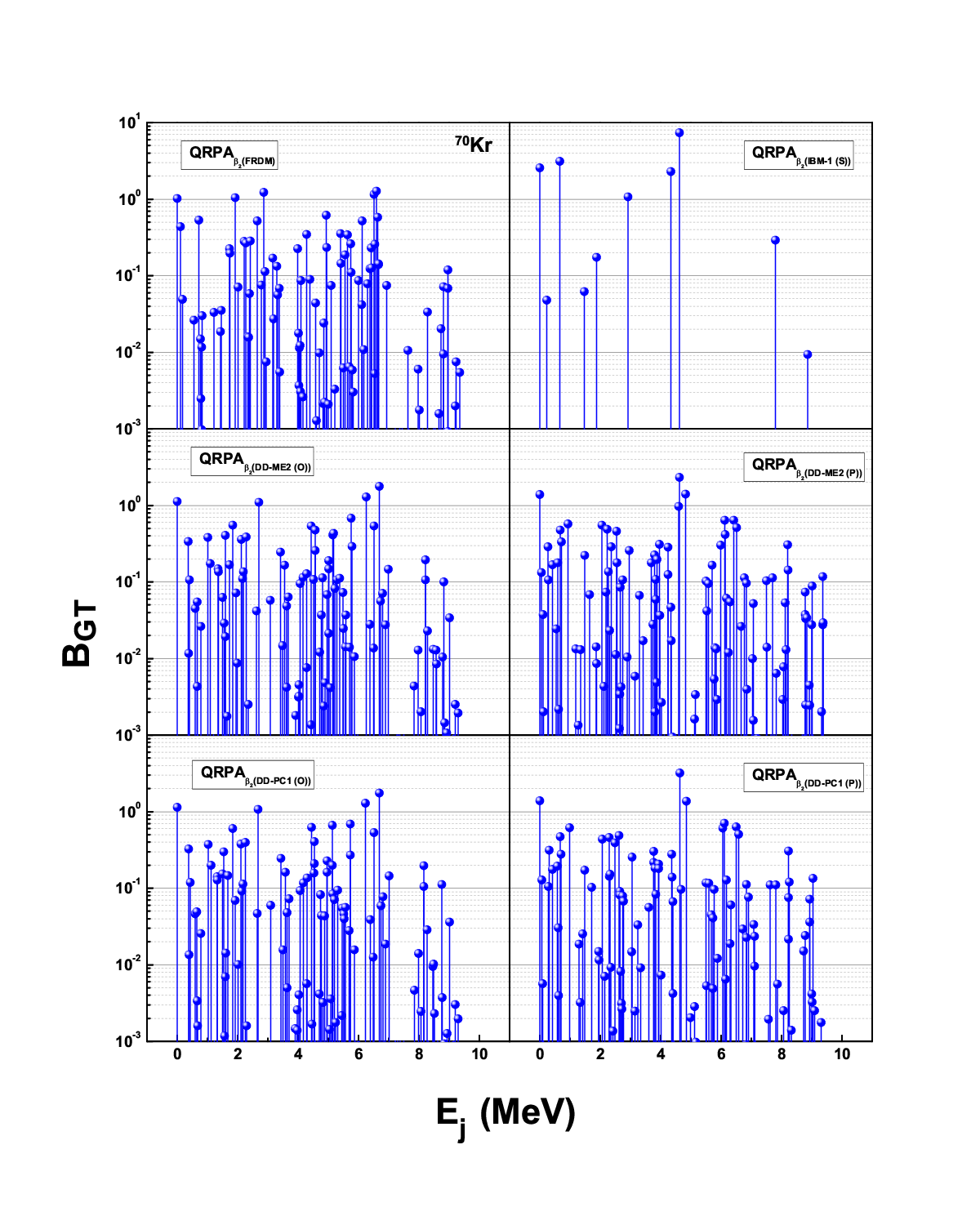}
	\vspace{-20mm}
	\caption{\centering Same as Fig.~\ref{F68Se} but for $^{70}$Kr.} \label{F70Kr}
\end{figure*}

\begin{figure*}[h]
	\centering
	\includegraphics[width=1\textwidth]{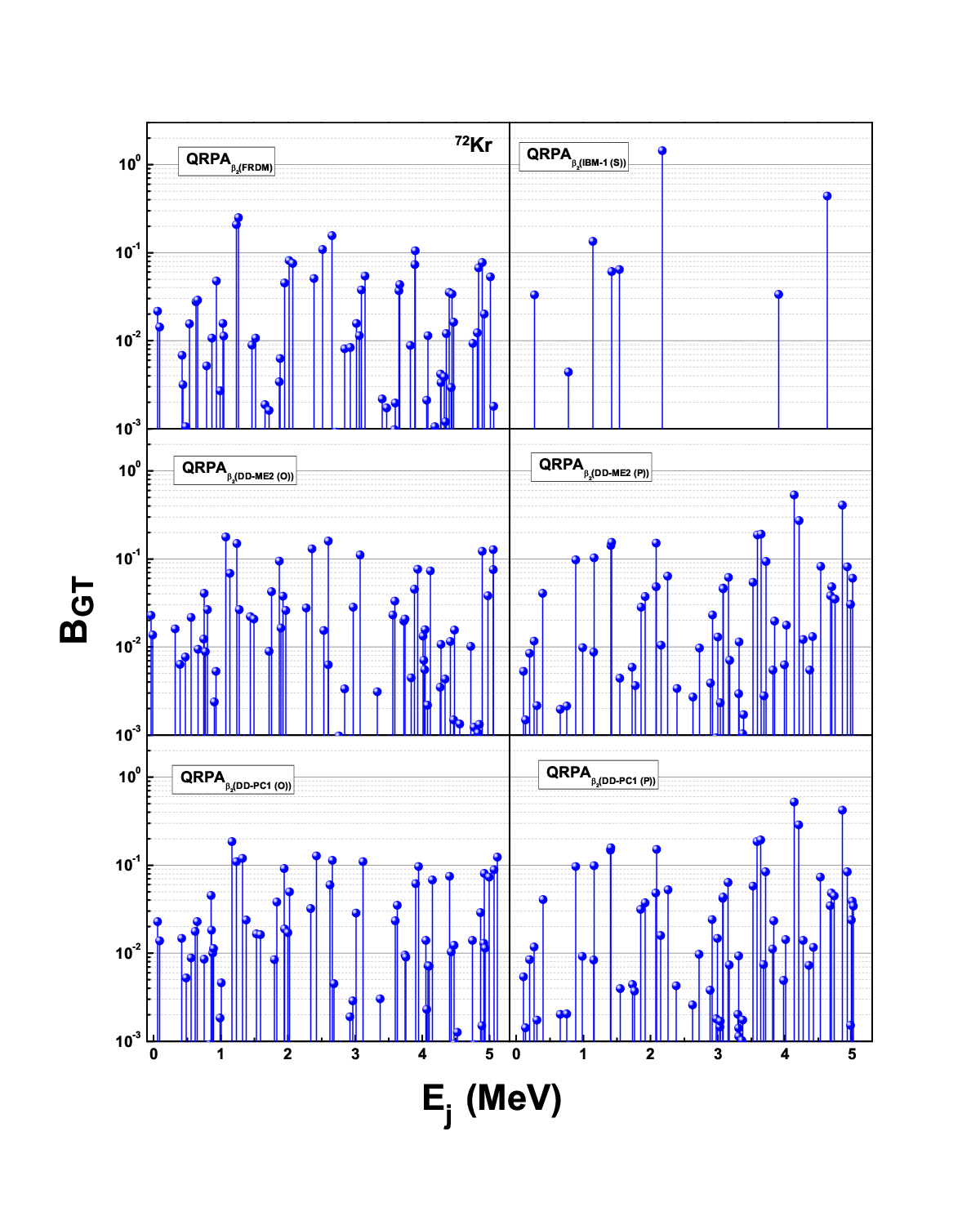}
	\vspace{-20mm}
	\caption{\centering Same as Fig.~\ref{F68Se} but for $^{72}$Kr.} \label{F72kr}
\end{figure*}

\begin{figure*}[h]
	\centering
	\includegraphics[width=1\textwidth]{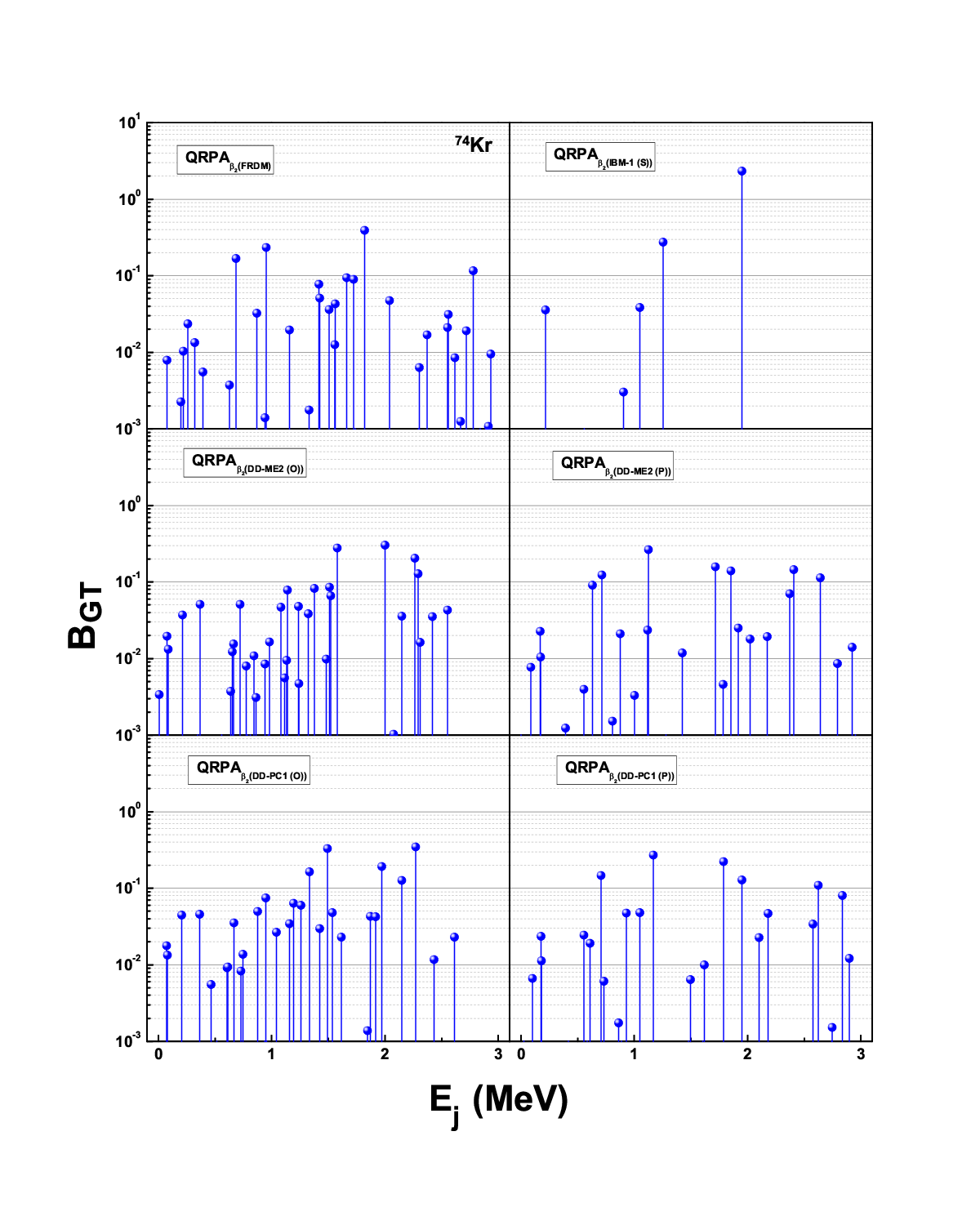}
	\vspace{-20mm}
	\caption{\centering Same as Fig.~\ref{F68Se} but for $^{74}$Kr.} \label{F74Kr}
\end{figure*}

\begin{figure*}[h]
	\centering
	\includegraphics[width=1\textwidth]{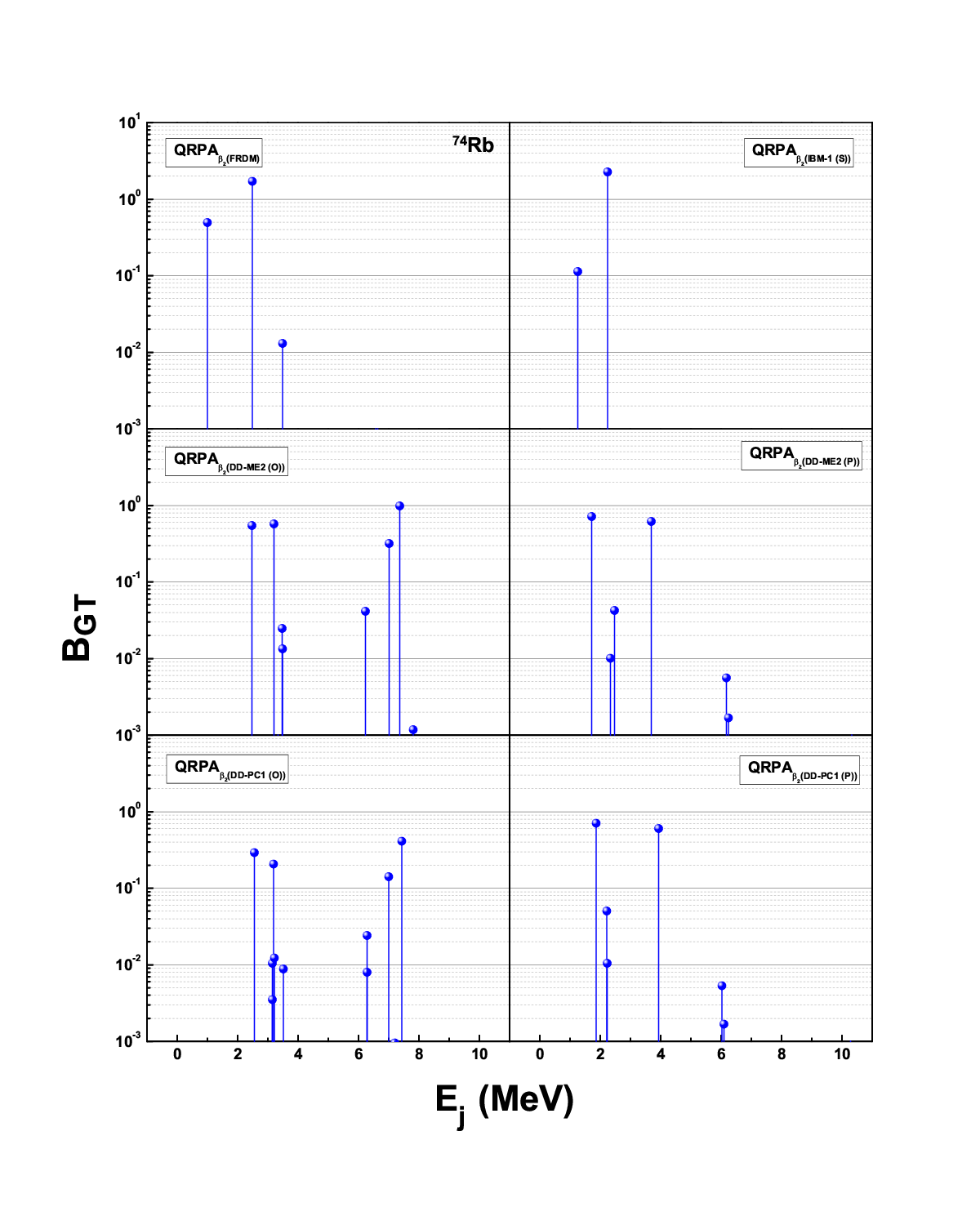}
	\vspace{-20mm}
	\caption{\centering Same as Fig.~\ref{F68Se} but for $^{74}$Rb.} \label{F74Rb}
\end{figure*}

\begin{figure*}[h]
	\centering
	\includegraphics[width=1\textwidth]{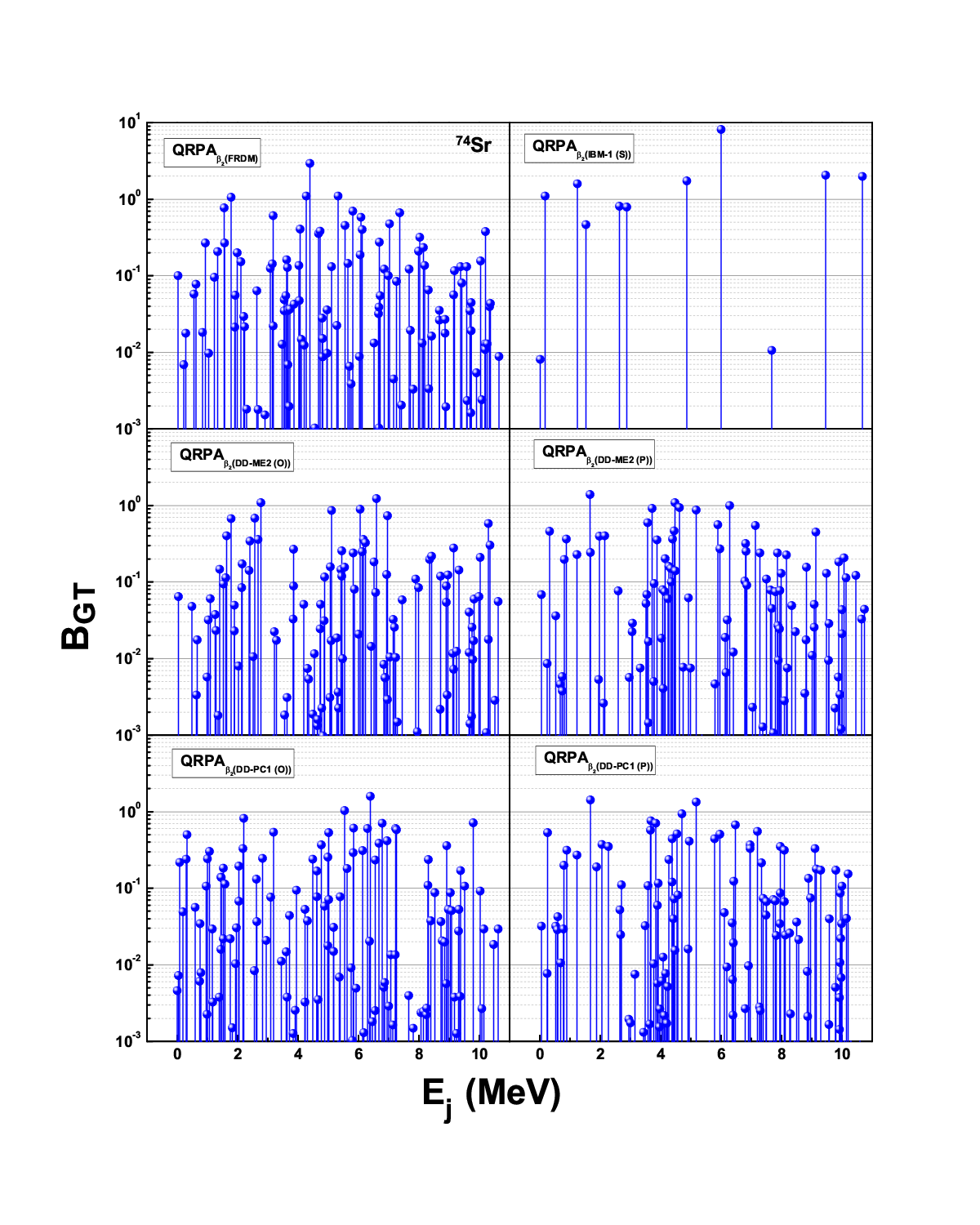}
	\vspace{-20mm}
	\caption{\centering Same as Fig.~\ref{F68Se} but for $^{74}$Sr.} \label{F74Sr}
\end{figure*}

Figures (\ref{F68Se} - \ref{F74Sr}) display the calculated GT strength distributions for the eight neutron-deficient nuclei. The abscissa shows the excitation energies in daughter nucleus and extends up to the Q$_{\beta^+}$ value. Each figure consists of six panels and the inset shows the type of interaction used to determine the $\beta_2$ values.  The strength distributions display  variations as $\beta_2$ values change. The calculated distributions remain almost unchanged for the QRPA$_{ \beta_2 (DD-PC1 (O))}$ \& QRPA$_{ \beta_2 (DD-ME2 (O))}$ and for the QRPA$_{ \beta_2 (DD-PC1 (P))}$ \& QRPA$_{ \beta_2 (DD-ME2 (P))}$ models, as the calculated $\beta_2$ values are quite close to each other. The QRPA$_{ \beta_2 (IBM-1 (S))}$ results predicted all nuclei to be spherical. This led to less fragmented strength distributions. Spherical deformation leads to concentration of most of the strength in few states \cite{hirsch1993}. 

Tables~(\ref{tab:68Se} - \ref{tab:74Sr}) present  the calculated state-by-state GT strength distributions, branching ratios, and partial half-lives as a function of $\beta_2$ values for $^{68}$Se, $^{70}$Se, $^{70}$Br, $^{70}$Kr, $^{72}$Kr, $^{74}$Kr, $^{74}$Rb \& $^{74}$Sr, respectively. The branching ratio (I) was computed using
\begin{eqnarray}
I = \frac{T_{1/2}}{t^p_{1/2}} \times 100 ~(\%).
\end{eqnarray}
Transitions possessing branching ratios greater than 1\% only are shown in Tables~(\ref{tab:68Se} - \ref{tab:74Sr}). It is again noted that the spherical deformation values led to smaller fragmentation of the GT strength in daughter states.

Tables~(\ref{Tab10} - \ref{Tab11}) show the calculated deformation parameters, total GT strength \& centroid values, measured  and calculated half-lives for selected nuclei.    The last column in the tables is a measure of the predictive power of the pn-QRPA model using different $\beta_2$ values.
The ratio $R_i$ was defined using
\begin{equation}
R_i = \left\{ \begin{array}{ll} T^{cal}_{1/2}/T^{exp}_{1/2} & \mbox{if $T^{cal}_{1/2}\geq T^{exp}_{1/2}$} \\ \\
	
	T^{exp}_{1/2}/T^{cal}_{1/2} & \mbox{if $T^{exp}_{1/2} > T^{cal}_{1/2}$}.
\end{array}
\label{Ri}
\right.
\end{equation} 
The nuclei selected for the current investigation find their respective positions far away from the $\beta$-stability line. It is noted from Tables~(\ref{Tab10} - \ref{Tab11})  that the FRDM and RMF models predicted large deformations for these nuclei.  The measured $\beta_2$ values were available only for four cases and suggested deformed shapes for these nuclei. Bigger total GT strength and smaller values of the centroid result in shorter computed $\beta$-decay half-lives.  For the case of $^{70}$Br and $^{74}$Rb, both the RMF and FRDM predicted $\beta_2$ values resulted in larger deviations in the calculated half-lives. Measured deformation values were not present for both these nuclei. For the case $^{74}$Rb, there is large deviation between the half-lives computed using QRPA$_{ \beta_2 (DD-PC1 (O))}$ \& QRPA$_{ \beta_2 (DD-ME2 (O))}$ interactions, even though the deformations are almost comparable. The reason is that the QRPA$_{ \beta_2 (DD-ME2 (O))}$ leads to higher computed cumulative strength and lower centroid value as compared to QRPA$_{ \beta_2 (DD-PC1 (O))}$ interaction, which led to smaller computed half life value. Tables~(\ref{Tab10} - \ref{Tab11}) show that the computed $R_i$ value for most of the cases is within  2, implying the model reproduces most of the experimentally known half-lives  within a factor of 2. Small $R_i$ values demonstrate the effectiveness of the pn-QRPA model. It was commented that the pn-QRPA model gives better prediction for nuclei far off from the line of stability \cite{hirsch1993,staudt1990}.

Table~\ref{Tab12} shows overall standings of the accuracy of the pn-QRPA model as a function of the computed  $\beta_2$ values. The average ratio ($\bar{R}$) was defined as
\begin{equation}
\bar{R} = \frac{\sum_{i=1}^{n}R_i}{n} ,
\label{Rbar}
\end{equation}
where $n$ is the number of nuclei.
The lowest ratio was reported for QRPA$_{ \beta_2 (NNDC)}$. However, it is remarked that this average was computed only for four out of the eight selected cases and may not be treated as a reliable data. Amongst the other models, the QRPA$_{ \beta_2 (DD-ME2 (O))}$ resulted in the smallest $\bar{R}$ value. This is closely followed by the QRPA$_{ \beta_2 (FRDM)}$ results. The QRPA$_{ \beta_2 (DD-PC1 (P))}$  and QRPA$_{ \beta_2 (IBM-1 (S))}$ predicted rather poor half-lives.

\section{Summary and Conclusion}
In this paper, we report the ground-state deformations and investigations on possible shape co-existence for  A $\sim$ 70 nuclei.

We calculated energy levels of even-even nuclei by fitting parameters of the IBM-1 Hamiltonian and obtained results in close proximity with the experimental data. Using the same set of parameters the {PESs} were plotted that predicted spherical shapes for the selected nuclei.

The RMF model with density-dependent meson exchange (DD-ME2) and point coupling (DD-PC1) interactions was later employed to compute binding and two-nucleon separation energies for the eight selected neutron-deficient nuclei.  The calculated quantities were in agreement with the measured data. The calculated PESs were used for determination of ground-state shape of the selected nuclei. Both (DD-ME2 and DD-PC1) functionals produced similar PESs and  predicted oblate configurations for $^{70}$Se, $^{70}$Br, $^{70}$Kr and $^{72}$Kr nuclei. A possible shape co-existence (oblate and prolate) was supported by the RMF model  for $^{68}$Se, $^{74}$Kr, $^{74}$Rb and $^{74}$Sr nucleus. On the other hand, the IBM-1 predicted \textit{zero} values for deformation parameters.

The pn-QRPA model with a schematic and separable potential was later employed to calculate $\beta$-decay properties of A $\sim$ 70 nuclei.  The different set of $\beta_2$ values including ones obtained from IBM-1 and RMF models were used as inputs to perform the systematic pn-QRPA calculations for the given neutron-deficient nucleus. The model has an excellent track-record for predicting half-lives of nuclei far-off from line of stability. Our investigation supported shape co-existence phenomena for neutron-deficient A $\sim$ 70 nuclei. The  predicted half-lives using QRPA$_{ \beta_2 (DD-ME2 (P))}$ \& QRPA$_{ \beta_2 (DD-ME2 (O))}$ were in best agreement with the measured data. Our findings may prove useful for further investigation of the $rp$-process waiting points.

\section*{Acknowledgments}
J.-U. Nabi would like to acknowledge the support of the Higher Education Commission Pakistan through project number 20-15394/NRPU/R$\&$D/HEC/2021.

%\begin{figure}[h]
%\centering
%\includegraphics[width=1\textwidth]{3_rprn.JPG}
%\caption{xxx.} \label{rprn}
%\end{figure}.

\clearpage
\newpage
\newpage

\newpage
\newpage

\newpage
\newpage
\newpage
\newpage
\newpage
\newpage
\newpage
\clearpage

\begin{table}
	\caption{Set of parameters of the IBM-1 Hamiltonian
		in units of keV. $N$ is the number of the bosons. $\chi$ is
		dimensionless.} \label{par} \centering
	\begin{tabular}{cccccc}
		\hline
		Nuclei & $N$ &$\epsilon$&$\kappa$&$\kappa'$&$\chi$\\
		\hline
		$^{68}$Se~&~6~&~544.2~&~39.9~&~32.5~&~-0.58~\\
		$^{70}$Se~&~7~&~470.2~&~79.4~&~27.0~&~-0.29~\\
		$^{70}$Kr~&~7~&~203.4~&~46.9~&~29.4~&~-0.90~\\
		$^{72}$Kr~&~8~&~203.4~&~46.9~&~29.4~&~-0.90~\\
		$^{74}$Kr~&~8~&~116.0~&~31.0~&~29.4~&~-0.50~\\
		$^{74}$Sr~&~9~&~728.9~&-20.5~&~-~&~-0.50~\\
		\hline
	\end{tabular}
\end{table}
\begin{table*}[htbp]
	\centering
	\caption{State-by-state $B_{GT}$ strength, branching ratios and partial half-lives calculated using the deformed pn-QRPA with six different deformation parameter values for $^{68}$Se.}
	\begin{adjustbox}{width=1\textwidth}
		\begin{tabular}{cccc|cccc|cccc}
			\hline\\
			\multicolumn{4}{c}{QRPA$_{ \beta_2 (FRDM)}$}       & \multicolumn{4}{c}{QRPA$_{ \beta_2 (IBM-1 (S))}$} & \multicolumn{4}{c}{QRPA$_{ \beta_2 (DD-ME2 (O))}$} \\
			\hline\\
			$E_j$ & $B_{GT}$   & I     & t$^p_{1/2}$ & $E_j$ & $B_{GT}$   & I     & t$^p_{1/2}$ & $E_j$ & $B_{GT}$   & I     & t$^p_{1/2}$ \\
			0.147 & 0.016 & 8.05  & 4.73E+02 & 0.103 & 0.025 & 4.09  & 2.81E+02 & 0.097 & 0.006 & 2.45  & 1.26E+03 \\
			0.222 & 0.003 & 1.52  & 2.51E+03 & 0.312 & 0.502 & 62.00 & 1.85E+01 & 0.347 & 0.118 & 37.27 & 8.26E+01 \\
			0.316 & 0.081 & 33.21 & 1.15E+02 & 0.489 & 0.032 & 3.14  & 3.66E+02 & 0.430 & 0.008 & 2.20  & 1.40E+03 \\
			0.608 & 0.007 & 1.94  & 1.97E+03 & 0.740 & 0.378 & 25.63 & 4.48E+01 & 0.575 & 0.078 & 18.06 & 1.70E+02 \\
			0.734 & 0.054 & 12.27 & 3.11E+02 & 1.328 & 0.194 & 5.03  & 2.28E+02 & 0.770 & 0.016 & 2.81  & 1.10E+03 \\
			0.766 & 0.020 & 4.33  & 8.81E+02 &       &       &       &          & 0.775 & 0.025 & 4.30  & 7.15E+02 \\
			0.828 & 0.091 & 17.90 & 2.13E+02 &       &       &       &          & 0.880 & 0.040 & 5.83  & 5.28E+02 \\
			0.904 & 0.017 & 3.03  & 1.26E+03 &       &       &       &          & 1.189 & 0.043 & 3.85  & 8.00E+02 \\
			1.161 & 0.020 & 2.33  & 1.63E+03 &       &       &       &          & 1.420 & 0.041 & 2.42  & 1.27E+03 \\
			1.360 & 0.049 & 3.94  & 9.67E+02 &       &       &       &          & 1.548 & 0.225 & 10.37 & 2.97E+02 \\
			1.426 & 0.015 & 1.09  & 3.51E+03 &       &       &       &          & 1.867 & 0.251 & 5.99  & 5.14E+02 \\
			1.843 & 0.051 & 1.60  & 2.39E+03 &       &       &       &          &       &       &       &          \\
			1.901 & 0.122 & 3.34  & 1.14E+03 &       &       &       &          &       &       &       &          \\
			1.961 & 0.080 & 1.92  & 1.98E+03 &       &       &       &          &       &       &       &         \\
			\hline\\
			\multicolumn{4}{c}{QRPA$_{ \beta_2 (DD-ME2 (P))}$}       & \multicolumn{4}{c}{QRPA$_{ \beta_2 (DD-PC1 (O))}$} & \multicolumn{4}{c}{QRPA$_{ \beta_2 (DD-PC1 (P))}$} \\
			\hline\\
			$E_j$ & $B_{GT}$   & I     & t$^p_{1/2}$ & $E_j$ & $B_{GT}$   & I     & t$^p_{1/2}$ & $E_j$ & $B_{GT}$   & I     & t$^p_{1/2}$ \\
			0.089 & 0.005 & 2.71  & 1.28E+03 & 0.096 & 0.006 & 2.47  & 1.24E+03 & 0.070 & 0.003 & 1.43  & 2.40E+03 \\
			0.231 & 0.084 & 34.83 & 9.92E+01 & 0.343 & 0.118 & 37.48 & 8.18E+01 & 0.082 & 0.003 & 1.66  & 2.06E+03 \\
			0.460 & 0.008 & 2.34  & 1.48E+03 & 0.428 & 0.008 & 2.18  & 1.41E+03 & 0.224 & 0.082 & 33.76 & 1.01E+02 \\
			0.667 & 0.155 & 35.13 & 9.83E+01 & 0.572 & 0.078 & 18.11 & 1.69E+02 & 0.445 & 0.008 & 2.33  & 1.46E+03 \\
			0.840 & 0.044 & 7.77  & 4.45E+02 & 0.771 & 0.017 & 2.92  & 1.05E+03 & 0.652 & 0.161 & 36.90 & 9.25E+01 \\
			0.893 & 0.010 & 1.64  & 2.11E+03 & 0.783 & 0.025 & 4.25  & 7.22E+02 & 0.845 & 0.030 & 5.14  & 6.64E+02 \\
			1.132 & 0.039 & 4.27  & 8.09E+02 & 0.877 & 0.041 & 6.00  & 5.11E+02 & 0.889 & 0.025 & 4.09  & 8.36E+02 \\
			1.447 & 0.060 & 3.75  & 9.21E+02 & 1.194 & 0.042 & 3.68  & 8.33E+02 & 1.138 & 0.038 & 4.05  & 8.42E+02 \\
			2.064 & 0.092 & 1.59  & 2.17E+03 & 1.425 & 0.035 & 2.01  & 1.52E+03 & 1.455 & 0.060 & 3.67  & 9.31E+02 \\
			2.164 & 0.115 & 1.57  & 2.21E+03 & 1.549 & 0.238 & 10.87 & 2.82E+02 & 2.081 & 0.086 & 1.42  & 2.41E+03 \\
			&       &       &          & 1.872 & 0.251 & 5.92  & 5.18E+02 & 2.179 & 0.112 & 1.45  & 2.35E+03\\
			\hline
		\end{tabular}%
	\end{adjustbox}
	\label{tab:68Se}%
\end{table*}%
%%%%%%%%%%%%%%%%%%%%%%%%%%%%%
%%%%%%%%%%%%%%%
\newpage
\begin{table*}[htbp]
	\centering
	\caption{Same as Table~\ref{tab:68Se} but for $^{70}$Se.}
	\begin{adjustbox}{width=1\textwidth}
		\begin{tabular}{cccc|cccc|cccc}
			\hline\\
			\multicolumn{4}{c}{QRPA$_{ \beta_2 (FRDM)}$}       & \multicolumn{4}{c}{QRPA$_{ \beta_2 (IBM-1 (S))}$} & \multicolumn{4}{c}{QRPA$_{ \beta_2 (DD-ME2 (O))}$} \\
			\hline\\
			$E_j$ & $B_{GT}$   & I     & t$^p_{1/2}$ & $E_j$ & $B_{GT}$   & I     & t$^p_{1/2}$ & $E_j$ & $B_{GT}$   & I     & t$^p_{1/2}$ \\
			0.059 & 0.013 & 8.89  & 3.18E+04 & 0.250 & 0.002 & 1.11  & 4.06E+05 & 0.028 & 0.001 & 1.06  & 2.60E+05 \\
			0.072 & 0.003 & 1.80  & 1.57E+05 & 0.790 & 0.393 & 73.86 & 6.12E+03 & 0.066 & 0.012 & 8.18  & 3.36E+04 \\
			0.173 & 0.003 & 1.46  & 1.94E+05 & 1.086 & 0.011 & 1.19  & 3.80E+05 & 0.077 & 0.003 & 1.65  & 1.67E+05 \\
			0.297 & 0.023 & 8.71  & 3.25E+04 & 1.417 & 0.037 & 2.25  & 2.01E+05 & 0.162 & 0.003 & 1.54  & 1.78E+05 \\
			0.517 & 0.055 & 11.86 & 2.39E+04 & 1.728 & 0.274 & 7.76  & 5.83E+04 & 0.301 & 0.039 & 14.06 & 1.96E+04 \\
			0.527 & 0.023 & 4.94  & 5.73E+04 & 1.982 & 1.279 & 13.83 & 3.27E+04 & 0.434 & 0.011 & 2.93  & 9.40E+04 \\
			0.619 & 0.038 & 6.48  & 4.36E+04 &       &       &       &          & 0.572 & 0.041 & 7.55  & 3.65E+04 \\
			0.735 & 0.139 & 18.30 & 1.55E+04 &       &       &       &          & 0.622 & 0.062 & 10.24 & 2.69E+04 \\
			1.137 & 0.234 & 14.84 & 1.91E+04 &       &       &       &          & 0.688 & 0.016 & 2.34  & 1.18E+05 \\
			1.192 & 0.312 & 18.04 & 1.57E+04 &       &       &       &          & 0.730 & 0.151 & 19.52 & 1.41E+04 \\
			&       &       &          &       &       &       &          & 0.903 & 0.017 & 1.58  & 1.74E+05 \\
			&       &       &          &       &       &       &          & 1.139 & 0.115 & 7.09  & 3.89E+04 \\
			&       &       &          &       &       &       &          & 1.182 & 0.040 & 2.30  & 1.20E+05 \\
			&       &       &          &       &       &       &          & 1.244 & 0.044 & 2.24  & 1.23E+05 \\
			&       &       &          &       &       &       &          & 1.266 & 0.206 & 10.21 & 2.70E+04 \\
			&       &       &          &       &       &       &          & 1.374 & 0.032 & 1.30  & 2.12E+05 \\
			&       &       &          &       &       &       &          & 1.376 & 0.048 & 1.95  & 1.41E+05\\
			\hline\\
			\multicolumn{4}{c}{QRPA$_{ \beta_2 (DD-ME2 (P))}$}       & \multicolumn{4}{c}{QRPA$_{ \beta_2 (DD-PC1 (O))}$} & \multicolumn{4}{c}{QRPA$_{ \beta_2 (DD-PC1 (P))}$} \\
			\hline\\
			$E_j$ & $B_{GT}$   & I     & t$^p_{1/2}$ & $E_j$ & $B_{GT}$   & I     & t$^p_{1/2}$ & $E_j$ & $B_{GT}$   & I     & t$^p_{1/2}$ \\
			0.037 & 0.005 & 4.75  & 7.16E+04 & 0.026 & 0.002 & 1.14  & 2.36E+05 & 0.037 & 0.005 & 4.75  & 7.16E+04 \\
			0.039 & 0.002 & 1.96  & 1.74E+05 & 0.062 & 0.012 & 7.79  & 3.47E+04 & 0.039 & 0.002 & 1.96  & 1.74E+05 \\
			0.095 & 0.002 & 1.19  & 2.85E+05 & 0.072 & 0.002 & 1.56  & 1.73E+05 & 0.095 & 0.002 & 1.19  & 2.85E+05 \\
			0.135 & 0.015 & 10.42 & 3.27E+04 & 0.152 & 0.003 & 1.54  & 1.75E+05 & 0.135 & 0.015 & 10.42 & 3.27E+04 \\
			0.201 & 0.007 & 4.26  & 7.98E+04 & 0.294 & 0.043 & 15.49 & 1.74E+04 & 0.201 & 0.007 & 4.26  & 7.98E+04 \\
			0.302 & 0.063 & 28.06 & 1.21E+04 & 0.434 & 0.018 & 4.45  & 6.07E+04 & 0.302 & 0.063 & 28.06 & 1.21E+04 \\
			0.369 & 0.063 & 23.83 & 1.43E+04 & 0.583 & 0.035 & 6.20  & 4.36E+04 & 0.369 & 0.063 & 23.83 & 1.43E+04 \\
			0.719 & 0.012 & 1.96  & 1.74E+05 & 0.622 & 0.062 & 9.99  & 2.70E+04 & 0.719 & 0.012 & 1.96  & 1.74E+05 \\
			0.837 & 0.037 & 4.79  & 7.10E+04 & 0.690 & 0.022 & 3.01  & 8.97E+04 & 0.837 & 0.037 & 4.79  & 7.10E+04 \\
			0.945 & 0.068 & 7.15  & 4.76E+04 & 0.724 & 0.152 & 19.54 & 1.38E+04 & 0.945 & 0.068 & 7.15  & 4.76E+04 \\
			1.232 & 0.029 & 1.87  & 1.82E+05 & 0.913 & 0.027 & 2.35  & 1.15E+05 & 1.232 & 0.029 & 1.87  & 1.82E+05 \\
			1.795 & 0.074 & 1.27  & 2.69E+05 & 1.096 & 0.043 & 2.82  & 9.59E+04 & 1.795 & 0.074 & 1.27  & 2.69E+05 \\
			1.815 & 0.313 & 5.05  & 6.74E+04 & 1.194 & 0.064 & 3.49  & 7.73E+04 & 1.815 & 0.313 & 5.05  & 6.74E+04 \\
			&       &       &          & 1.208 & 0.070 & 3.74  & 7.22E+04 &       &       &       &          \\
			&       &       &          & 1.262 & 0.052 & 2.56  & 1.06E+05 &       &       &       &          \\
			&       &       &          & 1.332 & 0.064 & 2.75  & 9.84E+04 &       &       &       &          \\
			&       &       &          & 1.337 & 0.158 & 6.74  & 4.01E+04 &       &       &       &         \\
			\hline
		\end{tabular}%
	\end{adjustbox}
	\label{tab:70Se}%
\end{table*}%
%%%%%%%%%%%%%%%%%%%%%%%%%%%%
%%%%%%%%%%%%%%%
\newpage
\begin{table*}[htbp]
	\centering
	\caption{Same as Table~\ref{tab:68Se} but for $^{70}$Br.}
	\begin{adjustbox}{width=1\textwidth}
		\begin{tabular}{cccc|cccc|cccc}
			\hline\\
			\multicolumn{4}{c}{QRPA$_{ \beta_2 (FRDM)}$}       & \multicolumn{4}{c}{QRPA$_{ \beta_2 (IBM-1 (S))}$} & \multicolumn{4}{c}{QRPA$_{ \beta_2 (DD-ME2 (O))}$} \\
			\hline\\
			$E_j$ & $B_{GT}$   & I     & t$^p_{1/2}$ & $E_j$ & $B_{GT}$   & I     & t$^p_{1/2}$ & $E_j$ & $B_{GT}$   & I     & t$^p_{1/2}$ \\
			1.259 & 0.279 & 41.18 & 2.76E-01 & 1.258 & 0.333 & 36.21 & 4.63E-01 & 1.237 & 0.099 & 16.95 & 7.66E-01 \\
			1.324 & 0.024 & 3.43  & 3.31E+00 & 2.329 & 1.103 & 62.11 & 2.70E-01 & 1.251 & 0.127 & 21.58 & 6.02E-01 \\
			2.878 & 0.022 & 1.32  & 8.60E+00 &       &       &       &          & 2.722 & 0.029 & 2.19  & 5.94E+00 \\
			3.244 & 0.224 & 10.94 & 1.04E+00 &       &       &       &          & 2.970 & 0.027 & 1.75  & 7.41E+00 \\
			3.338 & 0.113 & 5.18  & 2.19E+00 &       &       &       &          & 3.342 & 0.024 & 1.25  & 1.04E+01 \\
			3.745 & 0.030 & 1.05  & 1.08E+01 &       &       &       &          & 3.431 & 0.200 & 9.90  & 1.31E+00 \\
			3.842 & 0.036 & 1.20  & 9.48E+00 &       &       &       &          & 3.488 & 0.128 & 6.13  & 2.12E+00 \\
			4.719 & 0.621 & 10.93 & 1.04E+00 &       &       &       &          & 3.671 & 0.059 & 2.50  & 5.19E+00 \\
			5.171 & 0.451 & 5.55  & 2.05E+00 &       &       &       &          & 3.997 & 0.032 & 1.08  & 1.21E+01 \\
			5.406 & 0.577 & 5.81  & 1.95E+00 &       &       &       &          & 4.303 & 0.091 & 2.51  & 5.18E+00 \\
			5.683 & 1.170 & 9.24  & 1.23E+00 &       &       &       &          & 4.312 & 0.121 & 3.30  & 3.93E+00 \\
			&       &       &          &       &       &       &          & 5.018 & 0.400 & 6.36  & 2.04E+00 \\
			&       &       &          &       &       &       &          & 5.206 & 0.577 & 7.87  & 1.65E+00 \\
			&       &       &          &       &       &       &          & 5.592 & 1.168 & 11.44 & 1.14E+00\\
			\hline\\
			\multicolumn{4}{c}{QRPA$_{ \beta_2 (DD-ME2 (P))}$}       & \multicolumn{4}{c}{QRPA$_{ \beta_2 (DD-PC1 (O))}$} & \multicolumn{4}{c}{QRPA$_{ \beta_2 (DD-PC1 (P))}$} \\
			\hline\\
			$E_j$ & $B_{GT}$   & I     & t$^p_{1/2}$ & $E_j$ & $B_{GT}$   & I     & t$^p_{1/2}$ & $E_j$ & $B_{GT}$   & I     & t$^p_{1/2}$ \\
			1.398 & 0.013 & 1.94  & 1.30E+01 & 1.237 & 0.099 & 16.95 & 7.66E-01 & 1.393 & 0.012 & 2.16  & 1.38E+01 \\
			1.785 & 0.130 & 15.44 & 1.63E+00 & 1.251 & 0.127 & 21.58 & 6.02E-01 & 1.781 & 0.129 & 18.28 & 1.63E+00 \\
			1.785 & 0.130 & 15.44 & 1.63E+00 & 2.722 & 0.029 & 2.19  & 5.94E+00 & 1.781 & 0.129 & 18.28 & 1.63E+00 \\
			2.089 & 0.032 & 3.16  & 7.94E+00 & 2.970 & 0.027 & 1.75  & 7.41E+00 & 2.097 & 0.032 & 3.67  & 8.10E+00 \\
			2.101 & 0.011 & 1.05  & 2.38E+01 & 3.342 & 0.024 & 1.25  & 1.04E+01 & 2.097 & 0.011 & 1.23  & 2.43E+01 \\
			2.245 & 0.096 & 8.56  & 2.93E+00 & 3.431 & 0.100 & 4.95  & 2.62E+00 & 2.245 & 0.098 & 10.30 & 2.89E+00 \\
			3.223 & 0.158 & 7.11  & 3.53E+00 & 3.488 & 0.128 & 6.13  & 2.12E+00 & 3.196 & 0.162 & 8.87  & 3.36E+00 \\
			3.492 & 0.057 & 2.11  & 1.19E+01 & 3.671 & 0.059 & 2.50  & 5.19E+00 & 3.498 & 0.060 & 2.62  & 1.14E+01 \\
			3.492 & 0.057 & 2.11  & 1.19E+01 & 3.997 & 0.032 & 1.08  & 1.21E+01 & 3.498 & 0.060 & 2.62  & 1.14E+01 \\
			3.656 & 0.063 & 2.03  & 1.23E+01 & 4.303 & 0.091 & 2.51  & 5.18E+00 & 3.645 & 0.062 & 2.38  & 1.25E+01 \\
			3.734 & 0.301 & 9.13  & 2.75E+00 & 4.312 & 0.121 & 3.30  & 3.93E+00 & 3.908 & 0.045 & 1.40  & 2.13E+01 \\
			3.734 & 0.301 & 9.13  & 2.75E+00 & 5.018 & 0.400 & 6.36  & 2.04E+00 & 3.908 & 0.045 & 1.40  & 2.13E+01 \\
			3.991 & 0.241 & 5.93  & 4.24E+00 & 5.206 & 0.577 & 7.87  & 1.65E+00 & 3.968 & 0.248 & 7.36  & 4.05E+00 \\
			5.279 & 0.266 & 1.92  & 1.31E+01 & 5.592 & 1.168 & 11.44 & 1.14E+00 & 5.280 & 0.266 & 2.27  & 1.31E+01 \\
			5.283 & 0.360 & 2.59  & 9.70E+00 &       &       &       &          & 5.284 & 0.361 & 3.07  & 9.70E+00 \\
			5.848 & 0.322 & 1.20  & 2.09E+01 &       &       &       &          & 5.851 & 0.318 & 1.41  & 2.12E+01 \\
			&       &       &          &       &       &       &          & 5.861 & 0.306 & 1.34  & 2.23E+01 \\
			&       &       &          &       &       &       &          & 5.861 & 0.306 & 1.34  & 2.23E+01 \\
			\hline
		\end{tabular}%
	\end{adjustbox}
	\label{tab:70Br}%
\end{table*}%
%%%%%%%%%%%%%%%%%%%%%%%%%%%%
%%%%%%%%%%%%%%%
\newpage
\begin{table*}[htbp]
	\centering
	\caption{Same as Table~\ref{tab:68Se} but for $^{70}$Kr.}
	\begin{adjustbox}{width=1\textwidth}
		\begin{tabular}{cccc|cccc|cccc}
			\hline\\
			\multicolumn{4}{c}{QRPA$_{ \beta_2 (FRDM)}$}       & \multicolumn{4}{c}{QRPA$_{ \beta_2 (IBM-1 (S))}$} & \multicolumn{4}{c}{QRPA$_{ \beta_2 (DD-ME2 (O))}$} \\
			\hline\\
			$E_j$ & $B_{GT}$   & I     & t$^p_{1/2}$ & $E_j$ & $B_{GT}$   & I     & t$^p_{1/2}$ & $E_j$ & $B_{GT}$   & I     & t$^p_{1/2}$ \\
			0.000 & 1.016 & 34.09 & 1.43E-01 & 0.004 & 2.543 & 49.17 & 5.71E-02 & 0.000 & 1.129 & 37.60 & 1.28E-01 \\
			0.110 & 0.439 & 13.84 & 3.51E-01 & 0.663 & 3.109 & 40.78 & 6.88E-02 & 0.360 & 0.336 & 9.10  & 5.30E-01 \\
			0.170 & 0.049 & 1.49  & 3.26E+00 & 1.873 & 0.174 & 1.02  & 2.74E+00 & 0.411 & 0.106 & 2.79  & 1.73E+00 \\
			0.713 & 0.530 & 11.67 & 4.16E-01 & 2.918 & 1.075 & 2.79  & 1.01E+00 & 0.598 & 0.045 & 1.06  & 4.57E+00 \\
			1.735 & 0.225 & 2.53  & 1.92E+00 & 4.336 & 2.301 & 1.50  & 1.87E+00 & 0.663 & 0.055 & 1.24  & 3.88E+00 \\
			1.740 & 0.197 & 2.21  & 2.20E+00 & 4.622 & 7.367 & 3.45  & 8.14E-01 & 1.007 & 0.380 & 6.91  & 6.98E-01 \\
			1.924 & 1.045 & 10.25 & 4.74E-01 &       &       &       &          & 1.100 & 0.173 & 2.96  & 1.63E+00 \\
			2.218 & 0.281 & 2.22  & 2.19E+00 &       &       &       &          & 1.356 & 0.149 & 2.15  & 2.25E+00 \\
			2.264 & 0.266 & 2.03  & 2.39E+00 &       &       &       &          & 1.359 & 0.137 & 1.97  & 2.44E+00 \\
			2.417 & 0.283 & 1.92  & 2.54E+00 &       &       &       &          & 1.593 & 0.404 & 4.98  & 9.69E-01 \\
			2.646 & 0.520 & 2.94  & 1.66E+00 &       &       &       &          & 1.724 & 0.167 & 1.88  & 2.56E+00 \\
			2.869 & 1.223 & 5.73  & 8.48E-01 &       &       &       &          & 1.843 & 0.552 & 5.70  & 8.46E-01 \\
			&       &       &          &       &       &       &          & 2.127 & 0.359 & 3.01  & 1.60E+00 \\
			&       &       &          &       &       &       &          & 2.187 & 0.137 & 1.10  & 4.40E+00 \\
			&       &       &          &       &       &       &          & 2.285 & 0.385 & 2.87  & 1.68E+00 \\
			&       &       &          &       &       &       &          & 2.699 & 1.091 & 5.85  & 8.25E-01\\
			\hline\\
			\multicolumn{4}{c}{QRPA$_{ \beta_2 (DD-ME2 (P))}$}       & \multicolumn{4}{c}{QRPA$_{ \beta_2 (DD-PC1 (O))}$} & \multicolumn{4}{c}{QRPA$_{ \beta_2 (DD-PC1 (P))}$} \\
			\hline\\
			$E_j$ & $B_{GT}$   & I     & t$^p_{1/2}$ & $E_j$ & $B_{GT}$   & I     & t$^p_{1/2}$ & $E_j$ & $B_{GT}$   & I     & t$^p_{1/2}$ \\
			0.000 & 1.382 & 35.24 & 1.05E-01 & 0.000 & 1.135 & 37.69 & 1.28E-01 & 0.000 & 1.377 & 35.68 & 1.05E-01 \\
			0.059 & 0.132 & 3.26  & 1.13E+00 & 0.376 & 0.326 & 8.71  & 5.52E-01 & 0.063 & 0.128 & 3.20  & 1.17E+00 \\
			0.275 & 0.286 & 6.23  & 5.93E-01 & 0.427 & 0.119 & 3.08  & 1.56E+00 & 0.289 & 0.105 & 2.30  & 1.63E+00 \\
			0.288 & 0.106 & 2.29  & 1.61E+00 & 0.597 & 0.046 & 1.08  & 4.45E+00 & 0.315 & 0.311 & 6.73  & 5.58E-01 \\
			0.417 & 0.168 & 3.36  & 1.10E+00 & 0.651 & 0.049 & 1.11  & 4.33E+00 & 0.415 & 0.176 & 3.58  & 1.05E+00 \\
			0.590 & 0.177 & 3.20  & 1.16E+00 & 1.022 & 0.373 & 6.70  & 7.18E-01 & 0.585 & 0.193 & 3.56  & 1.05E+00 \\
			0.682 & 0.474 & 8.09  & 4.57E-01 & 1.118 & 0.198 & 3.35  & 1.44E+00 & 0.681 & 0.470 & 8.16  & 4.60E-01 \\
			0.719 & 0.334 & 5.58  & 6.62E-01 & 1.325 & 0.141 & 2.07  & 2.32E+00 & 0.696 & 0.277 & 4.76  & 7.89E-01 \\
			0.929 & 0.573 & 8.38  & 4.41E-01 & 1.328 & 0.128 & 1.89  & 2.55E+00 & 0.983 & 0.618 & 8.88  & 4.23E-01 \\
			1.490 & 0.222 & 2.25  & 1.64E+00 & 1.502 & 0.154 & 2.01  & 2.39E+00 & 1.482 & 0.173 & 1.79  & 2.10E+00 \\
			2.051 & 0.550 & 3.74  & 9.88E-01 & 1.543 & 0.298 & 3.79  & 1.27E+00 & 2.068 & 0.438 & 2.99  & 1.26E+00 \\
			2.233 & 0.488 & 2.89  & 1.28E+00 & 1.675 & 0.147 & 1.70  & 2.82E+00 & 2.302 & 0.461 & 2.64  & 1.42E+00 \\
			2.363 & 0.286 & 1.54  & 2.40E+00 & 1.832 & 0.597 & 6.20  & 7.76E-01 & 2.482 & 0.392 & 1.95  & 1.93E+00 \\
			2.538 & 0.458 & 2.14  & 1.73E+00 & 2.102 & 0.376 & 3.20  & 1.50E+00 & 2.619 & 0.488 & 2.17  & 1.73E+00 \\
			4.623 & 2.316 & 1.43  & 2.59E+00 & 2.253 & 0.394 & 3.00  & 1.61E+00 & 4.637 & 3.176 & 1.95  & 1.92E+00 \\
			&       &       &          & 2.682 & 1.075 & 5.83  & 8.26E-01 &       &       &       &         \\
			\hline
		\end{tabular}%
	\end{adjustbox}
	\label{tab:70Kr}%
\end{table*}%
%%%%%%%%%%%%%%%%%%%%%%%%%%%%
%%%%%%%%%%%%%%%
\newpage
\begin{table*}[htbp]
	\centering
	\caption{Same as Table~\ref{tab:68Se} but for $^{72}$Kr.}
	\begin{adjustbox}{width=1\textwidth}
		\begin{tabular}{cccc|cccc|cccc}
			\hline\\
			\multicolumn{4}{c}{QRPA$_{ \beta_2 (FRDM)}$}       & \multicolumn{4}{c}{QRPA$_{ \beta_2 (IBM-1 (S))}$} & \multicolumn{4}{c}{QRPA$_{ \beta_2 (DD-ME2 (O))}$} \\
			\hline\\
			$E_j$ & $B_{GT}$   & I     & t$^p_{1/2}$ & $E_j$ & $B_{GT}$   & I     & t$^p_{1/2}$ & $E_j$ & $B_{GT}$   & I     & t$^p_{1/2}$ \\
			0.061 & 0.022 & 9.40  & 1.95E+02 & 0.270 & 0.033 & 18.65 & 1.62E+02 & 0.061 & 0.023 & 10.09 & 1.84E+02 \\
			0.087 & 0.014 & 6.00  & 3.05E+02 & 0.774 & 0.004 & 1.32  & 2.28E+03 & 0.085 & 0.014 & 5.86  & 3.17E+02 \\
			0.425 & 0.007 & 1.93  & 9.50E+02 & 1.143 & 0.134 & 23.95 & 1.26E+02 & 0.415 & 0.016 & 4.67  & 3.98E+02 \\
			0.531 & 0.016 & 3.86  & 4.75E+02 & 1.418 & 0.061 & 7.12  & 4.24E+02 & 0.483 & 0.006 & 1.69  & 1.10E+03 \\
			0.631 & 0.028 & 6.09  & 3.01E+02 & 1.535 & 0.064 & 6.21  & 4.86E+02 & 0.562 & 0.008 & 1.87  & 9.93E+02 \\
			0.654 & 0.029 & 6.11  & 3.00E+02 & 2.175 & 1.442 & 42.68 & 7.07E+01 & 0.647 & 0.022 & 4.73  & 3.92E+02 \\
			0.868 & 0.011 & 1.72  & 1.07E+03 &       &       &       &          & 0.747 & 0.009 & 1.80  & 1.03E+03 \\
			0.933 & 0.048 & 6.96  & 2.63E+02 &       &       &       &          & 0.833 & 0.012 & 2.09  & 8.90E+02 \\
			1.030 & 0.016 & 1.99  & 9.19E+02 &       &       &       &          & 0.839 & 0.041 & 6.87  & 2.71E+02 \\
			1.046 & 0.011 & 1.41  & 1.30E+03 &       &       &       &          & 0.855 & 0.009 & 1.46  & 1.27E+03 \\
			1.232 & 0.208 & 19.74 & 9.28E+01 &       &       &       &          & 0.887 & 0.026 & 4.17  & 4.45E+02 \\
			1.261 & 0.249 & 22.54 & 8.13E+01 &       &       &       &          & 1.152 & 0.177 & 19.15 & 9.70E+01 \\
			1.951 & 0.045 & 1.27  & 1.45E+03 &       &       &       &          & 1.215 & 0.069 & 6.76  & 2.75E+02 \\
			2.019 & 0.082 & 2.00  & 9.14E+02 &       &       &       &          & 1.315 & 0.149 & 12.64 & 1.47E+02 \\
			2.070 & 0.076 & 1.68  & 1.09E+03 &       &       &       &          & 1.352 & 0.027 & 2.12  & 8.77E+02 \\
			&       &       &          &       &       &       &          & 1.510 & 0.022 & 1.36  & 1.36E+03 \\
			&       &       &          &       &       &       &          & 1.567 & 0.021 & 1.17  & 1.59E+03 \\
			&       &       &          &       &       &       &          & 1.819 & 0.042 & 1.53  & 1.21E+03 \\
			&       &       &          &       &       &       &          & 1.933 & 0.095 & 2.77  & 6.70E+02 \\
			&       &       &          &       &       &       &          & 2.410 & 0.130 & 1.44  & 1.29E+03 \\
			&       &       &          &       &       &       &          & 2.652 & 0.160 & 1.02  & 1.82E+03\\
			\hline\\
			\multicolumn{4}{c}{QRPA$_{ \beta_2 (DD-ME2 (P))}$}       & \multicolumn{4}{c}{QRPA$_{ \beta_2 (DD-PC1 (O))}$} & \multicolumn{4}{c}{QRPA$_{ \beta_2 (DD-PC1 (P))}$} \\
			\hline\\
			$E_j$ & $B_{GT}$   & I     & t$^p_{1/2}$ & $E_j$ & $B_{GT}$   & I     & t$^p_{1/2}$ & $E_j$ & $B_{GT}$   & I     & t$^p_{1/2}$ \\
			0.106 & 0.005 & 2.63  & 8.42E+02 & 0.061 & 0.023 & 9.69  & 1.85E+02 & 0.105 & 0.005 & 2.70  & 8.20E+02 \\
			0.200 & 0.008 & 3.80  & 5.83E+02 & 0.086 & 0.014 & 5.70  & 3.15E+02 & 0.201 & 0.008 & 3.79  & 5.85E+02 \\
			0.263 & 0.012 & 4.84  & 4.58E+02 & 0.418 & 0.015 & 4.13  & 4.35E+02 & 0.264 & 0.012 & 4.89  & 4.53E+02 \\
			0.395 & 0.041 & 14.42 & 1.54E+02 & 0.483 & 0.005 & 1.36  & 1.32E+03 & 0.396 & 0.041 & 14.40 & 1.54E+02 \\
			0.885 & 0.097 & 18.36 & 1.21E+02 & 0.558 & 0.009 & 2.07  & 8.67E+02 & 0.885 & 0.096 & 18.13 & 1.22E+02 \\
			0.986 & 0.010 & 1.62  & 1.37E+03 & 0.620 & 0.018 & 3.84  & 4.68E+02 & 0.985 & 0.009 & 1.51  & 1.47E+03 \\
			1.153 & 0.009 & 1.13  & 1.97E+03 & 0.647 & 0.023 & 4.81  & 3.74E+02 & 1.152 & 0.008 & 1.08  & 2.05E+03 \\
			1.158 & 0.102 & 13.13 & 1.69E+02 & 0.754 & 0.009 & 1.56  & 1.15E+03 & 1.156 & 0.098 & 12.63 & 1.75E+02 \\
			1.410 & 0.141 & 12.31 & 1.80E+02 & 0.853 & 0.045 & 7.26  & 2.47E+02 & 1.404 & 0.147 & 12.90 & 1.72E+02 \\
			1.418 & 0.154 & 13.25 & 1.67E+02 & 0.859 & 0.018 & 2.91  & 6.17E+02 & 1.410 & 0.157 & 13.64 & 1.62E+02 \\
			1.857 & 0.028 & 1.14  & 1.94E+03 & 0.879 & 0.010 & 1.55  & 1.16E+03 & 1.850 & 0.031 & 1.29  & 1.72E+03 \\
			1.919 & 0.037 & 1.34  & 1.66E+03 & 0.890 & 0.011 & 1.73  & 1.04E+03 & 1.917 & 0.037 & 1.34  & 1.65E+03 \\
			2.081 & 0.048 & 1.26  & 1.76E+03 & 1.165 & 0.184 & 18.95 & 9.48E+01 & 2.077 & 0.048 & 1.28  & 1.73E+03 \\
			2.084 & 0.152 & 3.97  & 5.59E+02 & 1.233 & 0.109 & 10.15 & 1.77E+02 & 2.088 & 0.151 & 3.92  & 5.65E+02 \\
			2.255 & 0.064 & 1.18  & 1.89E+03 & 1.326 & 0.119 & 9.56  & 1.88E+02 &       &       &       &          \\
			&       &       &          & 1.377 & 0.024 & 1.77  & 1.02E+03 &       &       &       &          \\
			&       &       &          & 1.832 & 0.038 & 1.31  & 1.37E+03 &       &       &       &          \\
			&       &       &          & 1.944 & 0.091 & 2.54  & 7.07E+02 &       &       &       &          \\
			&       &       &          & 2.024 & 0.050 & 1.19  & 1.51E+03 &       &       &       &          \\
			&       &       &          & 2.426 & 0.127 & 1.31  & 1.37E+03 &       &       &       &         \\
			\hline
		\end{tabular}%
	\end{adjustbox}
	\label{tab:72Kr}%
\end{table*}%
%%%%%%%%%%%%%%%%%%%%%%%%%%%%
%%%%%%%%%%%%%%%
\newpage
\begin{table*}[htbp]
	\centering
	\caption{Same as Table~\ref{tab:68Se} but for $^{74}$Kr.}
	\begin{adjustbox}{width=1\textwidth}
		\begin{tabular}{cccc|cccc|cccc}
			\hline\\
			\multicolumn{4}{c}{QRPA$_{ \beta_2 (FRDM)}$}       & \multicolumn{4}{c}{QRPA$_{ \beta_2 (IBM-1 (S))}$} & \multicolumn{4}{c}{QRPA$_{ \beta_2 (DD-ME2 (O))}$} \\
			\hline\\
			$E_j$ & $B_{GT}$   & I     & t$^p_{1/2}$ & $E_j$ & $B_{GT}$   & I     & t$^p_{1/2}$ & $E_j$ & $B_{GT}$   & I     & t$^p_{1/2}$ \\
			0.077 & 0.008 & 5.18  & 1.48E+04 & 0.216 & 0.036 & 27.40 & 4.42E+03 & 0.009 & 0.003 & 2.28  & 3.03E+04 \\
			0.199 & 0.002 & 1.13  & 6.78E+04 & 1.048 & 0.038 & 4.25  & 2.85E+04 & 0.077 & 0.019 & 11.49 & 6.02E+03 \\
			0.220 & 0.010 & 5.00  & 1.54E+04 & 1.256 & 0.273 & 19.47 & 6.22E+03 & 0.086 & 0.013 & 7.65  & 9.05E+03 \\
			0.262 & 0.024 & 10.38 & 7.39E+03 & 1.952 & 2.317 & 48.35 & 2.51E+03 & 0.212 & 0.037 & 16.34 & 4.24E+03 \\
			0.321 & 0.013 & 5.15  & 1.49E+04 &       &       &       &          & 0.369 & 0.051 & 15.94 & 4.34E+03 \\
			0.393 & 0.005 & 1.80  & 4.27E+04 &       &       &       &          & 0.652 & 0.012 & 1.98  & 3.50E+04 \\
			0.684 & 0.168 & 27.65 & 2.77E+03 &       &       &       &          & 0.664 & 0.016 & 2.42  & 2.86E+04 \\
			0.869 & 0.032 & 3.44  & 2.23E+04 &       &       &       &          & 0.723 & 0.051 & 6.93  & 9.99E+03 \\
			0.952 & 0.233 & 20.30 & 3.78E+03 &       &       &       &          & 0.844 & 0.011 & 1.10  & 6.31E+04 \\
			1.156 & 0.020 & 1.09  & 7.06E+04 &       &       &       &          & 0.982 & 0.016 & 1.20  & 5.75E+04 \\
			1.416 & 0.078 & 2.60  & 2.95E+04 &       &       &       &          & 1.082 & 0.047 & 2.75  & 2.51E+04 \\
			1.426 & 0.051 & 1.67  & 4.59E+04 &       &       &       &          & 1.139 & 0.078 & 4.04  & 1.71E+04 \\
			1.507 & 0.036 & 1.03  & 7.45E+04 &       &       &       &          & 1.238 & 0.048 & 2.01  & 3.44E+04 \\
			1.561 & 0.043 & 1.13  & 6.81E+04 &       &       &       &          & 1.322 & 0.038 & 1.38  & 5.03E+04 \\
			1.661 & 0.094 & 2.09  & 3.68E+04 &       &       &       &          & 1.377 & 0.083 & 2.68  & 2.59E+04 \\
			1.722 & 0.089 & 1.80  & 4.27E+04 &       &       &       &          & 1.512 & 0.085 & 2.18  & 3.18E+04 \\
			1.822 & 0.389 & 6.57  & 1.17E+04 &       &       &       &          & 1.521 & 0.066 & 1.66  & 4.17E+04 \\
			&       &       &          &       &       &       &          & 1.580 & 0.278 & 6.38  & 1.09E+04 \\
			&       &       &          &       &       &       &          & 2.002 & 0.302 & 3.24  & 2.13E+04 \\
			&       &       &          &       &       &       &          & 2.263 & 0.203 & 1.14  & 6.07E+04\\
			\hline\\
			\multicolumn{4}{c}{QRPA$_{ \beta_2 (DD-ME2 (P))}$}       & \multicolumn{4}{c}{QRPA$_{ \beta_2 (DD-PC1 (O))}$} & \multicolumn{4}{c}{QRPA$_{ \beta_2 (DD-PC1 (P))}$} \\
			\hline\\
			$E_j$ & $B_{GT}$   & I     & t$^p_{1/2}$ & $E_j$ & $B_{GT}$   & I     & t$^p_{1/2}$ & $E_j$ & $B_{GT}$   & I     & t$^p_{1/2}$ \\
			0.086 & 0.008 & 5.66  & 1.55E+04 & 0.072 & 0.018 & 9.76  & 6.53E+03 & 0.101 & 0.007 & 4.87  & 1.86E+04 \\
			0.169 & 0.023 & 13.90 & 6.31E+03 & 0.080 & 0.013 & 7.19  & 8.86E+03 & 0.178 & 0.023 & 14.59 & 6.20E+03 \\
			0.174 & 0.010 & 6.37  & 1.38E+04 & 0.206 & 0.044 & 18.38 & 3.47E+03 & 0.182 & 0.011 & 6.91  & 1.31E+04 \\
			0.555 & 0.004 & 1.02  & 8.63E+04 & 0.366 & 0.046 & 13.25 & 4.81E+03 & 0.555 & 0.024 & 6.48  & 1.39E+04 \\
			0.632 & 0.090 & 19.32 & 4.54E+03 & 0.464 & 0.005 & 1.27  & 5.02E+04 & 0.609 & 0.019 & 4.45  & 2.03E+04 \\
			0.714 & 0.124 & 21.76 & 4.03E+03 & 0.607 & 0.009 & 1.51  & 4.22E+04 & 0.707 & 0.146 & 26.86 & 3.37E+03 \\
			0.878 & 0.021 & 2.49  & 3.53E+04 & 0.613 & 0.009 & 1.51  & 4.22E+04 & 0.732 & 0.006 & 1.05  & 8.60E+04 \\
			1.119 & 0.023 & 1.60  & 5.48E+04 & 0.668 & 0.035 & 5.01  & 1.27E+04 & 0.932 & 0.047 & 5.07  & 1.78E+04 \\
			1.125 & 0.264 & 17.83 & 4.92E+03 & 0.729 & 0.008 & 1.03  & 6.22E+04 & 1.051 & 0.048 & 3.92  & 2.31E+04 \\
			1.715 & 0.158 & 3.67  & 2.39E+04 & 0.746 & 0.014 & 1.63  & 3.92E+04 & 1.169 & 0.270 & 17.15 & 5.27E+03 \\
			1.854 & 0.140 & 2.55  & 3.44E+04 & 0.877 & 0.050 & 4.29  & 1.49E+04 & 1.787 & 0.222 & 4.71  & 1.92E+04 \\
			&       &       &          & 0.950 & 0.075 & 5.43  & 1.17E+04 & 1.950 & 0.127 & 1.99  & 4.55E+04 \\
			&       &       &          & 1.042 & 0.026 & 1.56  & 4.09E+04 &       &       &       &          \\
			&       &       &          & 1.158 & 0.034 & 1.57  & 4.06E+04 &       &       &       &          \\
			&       &       &          & 1.194 & 0.063 & 2.70  & 2.37E+04 &       &       &       &          \\
			&       &       &          & 1.260 & 0.059 & 2.21  & 2.88E+04 &       &       &       &          \\
			&       &       &          & 1.336 & 0.164 & 5.29  & 1.21E+04 &       &       &       &          \\
			&       &       &          & 1.494 & 0.330 & 8.04  & 7.93E+03 &       &       &       &          \\
			&       &       &          & 1.536 & 0.048 & 1.09  & 5.86E+04 &       &       &       &          \\
			&       &       &          & 1.971 & 0.192 & 2.03  & 3.14E+04 &       &       &       &          \\
			&       &       &          & 2.271 & 0.344 & 1.74  & 3.66E+04 &       &       &       &         \\
			\hline
		\end{tabular}%
	\end{adjustbox}
	\label{tab:74Kr}%
\end{table*}%
%%%%%%%%%%%%%%%%%%%%%%%%%%%%
%%%%%%%%%%%%%%%
\newpage
\begin{table}[htbp]
	\centering
	\caption{Same as Table~\ref{tab:68Se} but for $^{74}$Rb.}
	\begin{adjustbox}{width=1\textwidth}
		\begin{tabular}{cccc|cccc|cccc}
			\hline\\
			\multicolumn{4}{c}{QRPA$_{ \beta_2 (FRDM)}$}       & \multicolumn{4}{c}{QRPA$_{ \beta_2 (IBM-1 (S))}$} & \multicolumn{4}{c}{QRPA$_{ \beta_2 (DD-ME2 (O))}$} \\
			\hline\\
			$E_j$ & $B_{GT}$   & I     & t$^p_{1/2}$ & $E_j$ & $B_{GT}$   & I     & t$^p_{1/2}$ & $E_j$ & $B_{GT}$   & I     & t$^p_{1/2}$ \\
			1.002 & 0.493 & 41.77 & 2.92E-01 & 1.262 & 0.113 & 8.43  & 1.47E+00 & 2.478 & 0.541 & 57.56 & 3.96E-01 \\
			2.480 & 1.703 & 58.02 & 2.11E-01 & 2.247 & 2.261 & 91.57 & 1.36E-01 & 3.198 & 0.573 & 38.45 & 5.92E-01 \\
			&       &       &          &       &       &       &          & 3.478 & 0.025 & 1.37  & 1.66E+01 \\
			&       &       &          &       &       &       &          & 7.368 & 0.976 & 1.07  & 2.12E+01\\
			\hline\\
			\multicolumn{4}{c}{QRPA$_{ \beta_2 (DD-ME2 (P))}$}       & \multicolumn{4}{c}{QRPA$_{ \beta_2 (DD-PC1 (O))}$} & \multicolumn{4}{c}{QRPA$_{ \beta_2 (DD-PC1 (P))}$} \\
			\hline\\
			$E_j$ & $B_{GT}$   & I     & t$^p_{1/2}$ & $E_j$ & $B_{GT}$   & I     & t$^p_{1/2}$ & $E_j$ & $B_{GT}$   & I     & t$^p_{1/2}$ \\
			1.713 & 0.717 & 79.32 & 3.05E-01 & 2.559 & 0.291 & 47.12 & 7.72E-01 & 1.864 & 0.704 & 79.30 & 3.41E-01 \\
			2.470 & 0.042 & 2.86  & 8.46E+00 & 3.146 & 0.010 & 1.16  & 3.13E+01 & 2.213 & 0.050 & 4.53  & 5.98E+00 \\
			3.686 & 0.617 & 17.06 & 1.42E+00 & 3.187 & 0.207 & 22.41 & 1.62E+00 & 3.928 & 0.601 & 15.22 & 1.78E+00 \\
			&       &       &          & 3.187 & 0.207 & 22.41 & 1.62E+00 &       &       &       &          \\
			&       &       &          & 3.214 & 0.012 & 1.31  & 2.79E+01 &       &       &       &          \\
			&       &       &          & 3.214 & 0.012 & 1.31  & 2.79E+01 &       &       &       &         \\
			\hline
		\end{tabular}%
	\end{adjustbox}
	\label{tab:74Rb}%
\end{table}%
%%%%%%%%%%%%%%
\newpage
\begin{table*}[htbp]
	\centering
	\caption{Same as Table~\ref{tab:68Se} but for $^{74}$Sr.}
	\begin{adjustbox}{width=1\textwidth}
		\begin{tabular}{cccc|cccc|cccc}
			\hline\\
			\multicolumn{4}{c}{QRPA$_{ \beta_2 (FRDM)}$}       & \multicolumn{4}{c}{QRPA$_{ \beta_2 (IBM-1 (S))}$} & \multicolumn{4}{c}{QRPA$_{ \beta_2 (DD-ME2 (O))}$} \\
			\hline\\
			$E_j$ & $B_{GT}$   & I     & t$^p_{1/2}$ & $E_j$ & $B_{GT}$   & I     & t$^p_{1/2}$ & $E_j$ & $B_{GT}$   & I     & t$^p_{1/2}$ \\
			0.024 & 0.100 & 4.47  & 7.22E-01 & 0.176 & 1.099 & 39.47 & 7.08E-02 & 0.040 & 0.064 & 3.84  & 1.14E+00 \\
			0.548 & 0.058 & 1.98  & 1.63E+00 & 1.247 & 1.593 & 32.60 & 8.57E-02 & 0.490 & 0.048 & 2.28  & 1.91E+00 \\
			0.615 & 0.077 & 2.56  & 1.26E+00 & 1.526 & 0.463 & 8.10  & 3.45E-01 & 1.029 & 0.032 & 1.14  & 3.82E+00 \\
			0.926 & 0.269 & 7.57  & 4.27E-01 & 2.636 & 0.809 & 7.16  & 3.90E-01 & 1.099 & 0.060 & 2.09  & 2.09E+00 \\
			1.232 & 0.095 & 2.27  & 1.42E+00 & 2.883 & 0.786 & 5.89  & 4.74E-01 & 1.257 & 0.037 & 1.19  & 3.68E+00 \\
			1.342 & 0.206 & 4.61  & 7.00E-01 & 4.859 & 1.719 & 2.68  & 1.04E+00 & 1.411 & 0.147 & 4.28  & 1.02E+00 \\
			1.560 & 0.769 & 15.24 & 2.12E-01 & 6.001 & 8.068 & 3.79  & 7.38E-01 & 1.524 & 0.094 & 2.58  & 1.69E+00 \\
			1.563 & 0.268 & 5.31  & 6.08E-01 &       &       &       &          & 1.605 & 0.114 & 2.96  & 1.47E+00 \\
			1.779 & 1.054 & 18.38 & 1.76E-01 &       &       &       &          & 1.631 & 0.403 & 10.35 & 4.22E-01 \\
			1.985 & 0.199 & 3.07  & 1.05E+00 &       &       &       &          & 1.778 & 0.672 & 15.85 & 2.75E-01 \\
			2.103 & 0.152 & 2.19  & 1.48E+00 &       &       &       &          & 1.886 & 0.050 & 1.10  & 3.96E+00 \\
			3.148 & 0.144 & 1.04  & 3.12E+00 &       &       &       &          & 2.129 & 0.084 & 1.61  & 2.71E+00 \\
			3.169 & 0.611 & 4.34  & 7.44E-01 &       &       &       &          & 2.153 & 0.172 & 3.23  & 1.35E+00 \\
			4.067 & 0.405 & 1.46  & 2.22E+00 &       &       &       &          & 2.379 & 0.141 & 2.29  & 1.90E+00 \\
			4.268 & 1.100 & 3.35  & 9.64E-01 &       &       &       &          & 2.410 & 0.342 & 5.46  & 7.98E-01 \\
			4.395 & 2.931 & 8.01  & 4.03E-01 &       &       &       &          & 2.568 & 0.675 & 9.75  & 4.47E-01 \\
			5.319 & 1.095 & 1.26  & 2.56E+00 &       &       &       &          & 2.679 & 0.361 & 4.84  & 9.01E-01 \\
			&       &       &          &       &       &       &          & 2.764 & 1.087 & 13.79 & 3.16E-01 \\
			&       &       &          &       &       &       &          & 3.858 & 0.266 & 1.52  & 2.86E+00 \\
			&       &       &          &       &       &       &          & 5.110 & 0.853 & 1.63  & 2.67E+00\\
			\hline\\
			\multicolumn{4}{c}{QRPA$_{ \beta_2 (DD-ME2 (P))}$}       & \multicolumn{4}{c}{QRPA$_{ \beta_2 (DD-PC1 (O))}$} & \multicolumn{4}{c}{QRPA$_{ \beta_2 (DD-PC1 (P))}$} \\
			\hline\\
			$E_j$ & $B_{GT}$   & I     & t$^p_{1/2}$ & $E_j$ & $B_{GT}$   & I     & t$^p_{1/2}$ & $E_j$ & $B_{GT}$   & I     & t$^p_{1/2}$ \\
			0.051 & 0.068 & 2.72  & 1.07E+00 & 0.075 & 0.216 & 8.41  & 3.43E-01 & 0.060 & 0.032 & 1.26  & 2.31E+00 \\
			0.329 & 0.459 & 15.90 & 1.83E-01 & 0.185 & 0.049 & 1.81  & 1.60E+00 & 0.261 & 0.529 & 18.93 & 1.53E-01 \\
			0.523 & 0.036 & 1.14  & 2.56E+00 & 0.291 & 0.239 & 8.38  & 3.44E-01 & 0.597 & 0.042 & 1.28  & 2.27E+00 \\
			0.807 & 0.196 & 5.32  & 5.47E-01 & 0.322 & 0.498 & 17.17 & 1.68E-01 & 0.796 & 0.199 & 5.40  & 5.38E-01 \\
			0.882 & 0.363 & 9.44  & 3.09E-01 & 0.596 & 0.056 & 1.68  & 1.72E+00 & 0.895 & 0.313 & 8.07  & 3.60E-01 \\
			1.234 & 0.226 & 4.87  & 5.99E-01 & 0.963 & 0.106 & 2.63  & 1.10E+00 & 1.233 & 0.271 & 5.80  & 5.01E-01 \\
			1.666 & 1.389 & 23.35 & 1.25E-01 & 0.995 & 0.243 & 5.90  & 4.89E-01 & 1.673 & 1.421 & 23.73 & 1.22E-01 \\
			1.673 & 0.243 & 4.07  & 7.16E-01 & 1.067 & 0.301 & 7.01  & 4.11E-01 & 1.875 & 0.190 & 2.81  & 1.03E+00 \\
			1.973 & 0.398 & 5.57  & 5.23E-01 & 1.442 & 0.140 & 2.65  & 1.09E+00 & 2.055 & 0.371 & 4.92  & 5.90E-01 \\
			2.145 & 0.402 & 5.07  & 5.74E-01 & 1.461 & 0.138 & 2.59  & 1.11E+00 & 2.264 & 0.351 & 4.10  & 7.09E-01 \\
			3.565 & 0.595 & 2.86  & 1.02E+00 & 1.522 & 0.183 & 3.30  & 8.73E-01 & 3.664 & 0.572 & 2.54  & 1.14E+00 \\
			3.719 & 0.917 & 3.92  & 7.44E-01 & 1.586 & 0.113 & 1.97  & 1.47E+00 & 3.679 & 0.757 & 3.33  & 8.73E-01 \\
			3.880 & 0.354 & 1.33  & 2.19E+00 & 2.037 & 0.194 & 2.59  & 1.11E+00 & 3.835 & 0.702 & 2.73  & 1.06E+00 \\
			4.451 & 0.466 & 1.10  & 2.66E+00 & 2.175 & 0.329 & 4.03  & 7.16E-01 & 4.385 & 0.439 & 1.09  & 2.67E+00 \\
			4.468 & 1.087 & 2.52  & 1.16E+00 & 2.208 & 0.818 & 9.83  & 2.93E-01 & 4.542 & 0.511 & 1.11  & 2.63E+00 \\
			4.617 & 0.937 & 1.90  & 1.53E+00 & 2.628 & 0.131 & 1.20  & 2.40E+00 & 4.705 & 0.938 & 1.75  & 1.66E+00 \\
			5.171 & 0.867 & 1.04  & 2.79E+00 & 2.831 & 0.246 & 1.97  & 1.46E+00 & 5.181 & 1.327 & 1.58  & 1.84E+00 \\
			&       &       &          & 3.187 & 0.539 & 3.38  & 8.55E-01 &       &       &       &         \\
			\hline
		\end{tabular}%
	\end{adjustbox}
	\label{tab:74Sr}%
\end{table*}%
%%%%%%%%%%%%%%%

\begin{table*}[]
	\tiny
	\centering
	%\vskip-80pt
	\caption{Computed deformation parameters ($\beta_2$),  total strength ($\sum$ $B_{GT}$), centroid ($\bar{E}$), measured and calculated half-lives for $^{68, 70}$Se, $^{70}$Br and $^{70}$Kr. $R_i$ is a measure of the predictive power of the model and described in Eq.~(\ref{Ri}).} \label{Tab10}
	% \begin{adjustbox}{width=1\textwidth}

		\begin{tabular}{cccccccccc}
			\hline\\
			Nuclei & Decay mode& Q$_{\beta^+}$ & Models & $ \beta_{2}$ &$\sum {B_{GT}}$& $\bar{E}$ &$T_{1/2}^{pnQRPA}$ (s)& $T_{1/2}^{Exp}$ (s)  & $R_i$\\
			&  & ($MeV$) & & &  & ($MeV$)&\\
			\hline\\
			$^{68}$Se & $\beta^+$ & 4.71  & QRPA$_{ \beta_2 (FRDM)}$  & 0.23300  & 3.18  & 3.43 & 3.81E+01 & 3.55E+01 ($\pm$ 0.7) & 1.07   \\
			&           &       & QRPA$_{ \beta_2 (NNDC)}$  & 0.24200  & 3.14  & 3.47 & 3.97E+01 &          & 1.12   \\
			&           &       & QRPA$_{ \beta_2 (IBM-1 (S))}$ & 0.00000  & 4.10  & 3.23 & 1.15E+01 &          & 3.09   \\
			&           &       & QRPA$_{ \beta_2 (DD-ME2 (O))}$ & -0.26900 & 2.78  & 3.06 & 3.08E+01 &          & 1.15   \\
			&           &       & QRPA$_{ \beta_2 (DD-ME2 (P))}$ & 0.25100  & 3.19  & 3.42 & 3.45E+01 &          & 1.03   \\
			&           &       & QRPA$_{ \beta_2 (DD-PC1 (O))}$ & -0.27000 & 2.67  & 3.04 & 3.07E+01 &          & 1.16   \\
			&           &       & QRPA$_{ \beta_2 (DD-PC1 (P))}$ & 0.25300  & 3.21  & 3.42 & 3.41E+01 &          & 1.04   \\
			&           &       &       &          &       &      &          &          &        \\
			$^{70}$Se & $\beta^+$ & 2.40  & QRPA$_{ \beta_2 (FRDM)}$  & -0.30700 & 1.16  & 1.24 & 2.83E+03 & 2.47E+03 ($\pm$ 18)& 1.15   \\
			&           &       & QRPA$_{ \beta_2 (NNDC)}$  & 0.20700  & 1.05  & 1.51 & 3.27E+03 &          & 1.33   \\
			&           &       & QRPA$_{ \beta_2 (IBM-1 (S))}$ & 0.00000  & 2.00  & 1.70 & 4.52E+03 &          & 1.83   \\
			&           &       & QRPA$_{ \beta_2 (DD-ME2 (O))}$ & -0.28100 & 1.16  & 1.24 & 2.75E+03 &          & 1.12   \\
			&           &       & QRPA$_{ \beta_2 (DD-ME2 (P))}$ & 0.23000  & 1.21  & 1.65 & 3.40E+03 &          & 1.38   \\
			&           &       & QRPA$_{ \beta_2 (DD-PC1 (O))}$ & -0.27300 & 1.15  & 1.23 & 2.70E+03 &          & 1.10   \\
			&           &       & QRPA$_{ \beta_2 (DD-PC1 (P))}$ & 0.23000  & 1.21  & 1.65 & 3.40E+03 &          & 1.38   \\
			&           &       &       &          &       &      &          &          &        \\
			$^{70}$Br & $\beta^+$ & 10.50 & QRPA$_{ \beta_2 (FRDM)}$  & -0.32700 & 10.25 & 7.44 & 1.14E-01 & 7.88E-02 ($\pm$ 0.0003)& 3.73   \\
			&           &       & QRPA$_{ \beta_2 (IBM-1 (S))}$ & 0.00000  & 12.77 & 7.30 & 1.68E-01 &          & 2.13   \\
			&           &       & QRPA$_{ \beta_2 (DD-ME2 (O))}$ & -0.30800 & 8.93  & 7.16 & 1.30E-01 &          & 1.65   \\
			&           &       & QRPA$_{ \beta_2 (DD-ME2 (P))}$ & 0.21400  & 12.24 & 7.31 & 2.51E-01 &          & 3.19   \\
			&           &       & QRPA$_{ \beta_2 (DD-PC1 (O))}$ & -0.30800 & 8.93  & 7.16 & 1.30E-01 &          & 1.65   \\
			&           &       & QRPA$_{ \beta_2 (DD-PC1 (P))}$ & 0.21500  & 11.76 & 7.46 & 2.98E-01 &          & 3.78   \\
			&           &       &       &          &       &      &          &          &        \\
			$^{70}$Kr & $\beta^+$ & 9.38  & QRPA$_{ \beta_2 (FRDM)}$  & -0.32700 & 15.49 & 4.09 & 4.86E-02 & 4.50E-02 ($\pm$ 0.00014)& 1.08   \\
			&           &       & QRPA$_{ \beta_2 (IBM-1 (S))}$ & 0.00000  & 16.98 & 3.06 & 2.81E-02 &          & 1.60   \\
			&           &       & QRPA$_{ \beta_2 (DD-ME2 (O))}$ & -0.28800 & 15.66 & 4.09 & 4.82E-02 &          & 1.07   \\
			&           &       & QRPA$_{ \beta_2 (DD-ME2 (P))}$ & 0.24100  & 17.76 & 3.86 & 3.69E-02 &          & 1.22   \\
			&           &       & QRPA$_{ \beta_2 (DD-PC1 (O))}$ & -0.28500 & 15.67 & 4.08 & 4.81E-02 &          & 1.07   \\
			&           &       & QRPA$_{ \beta_2 (DD-PC1 (P))}$ & 0.24700  & 17.62 & 3.84 & 3.75E-02 &          & 1.20   \\
			&           &       &       &          &       &      &          &          &        \\
			\hline
		\end{tabular}
		%\end{adjustbox}
	\end{table*}
	\clearpage
	\begin{table*}[]
		\tiny
		\centering
		%\vskip-80pt
		\caption{\centering Same as Table \ref{Tab10}, but for $^{72, 74}$Kr, $^{74}$Rb and $^{74}$Sr.} \label{Tab11}
		% \begin{adjustbox}{width=1\textwidth}
			\begin{tabular}{cccccccccc}
				\hline\\
				Nuclei & Decay mode& Q$_{\beta^+}$ & Models & $ \beta_{2}$ &$\sum {B_{GT}}$& $\bar{E}$ &$T_{1/2}^{pnQRPA}$ (s)& $T_{1/2}^{Exp}$ (s)  & $R_i$\\
				&  &  ($MeV$)& & &  & ($MeV$)&\\
				\hline\\
				$^{72}$Kr & $\beta^+$ & 5.12  & QRPA$_{ \beta_2 (FRDM)}$  & -0.36600 & 2.00  & 2.60 & 1.83E+01 & 1.71E+01 ($\pm$ 0.18)& 1.07   \\
				&           &       & QRPA$_{ \beta_2 (NNDC)}$  & 0.33200  & 3.37  & 3.38 & 2.13E+01 &          & 1.25   \\
				&           &       & QRPA$_{ \beta_2 (IBM-1 (S))}$ & 0.00000  & 2.21  & 2.56 & 3.02E+01 &          & 1.76   \\
				&           &       & QRPA$_{ \beta_2 (DD-ME2 (O))}$ & -0.34600 & 2.14  & 2.75 & 1.86E+01 &          & 1.09   \\
				&           &       & QRPA$_{ \beta_2 (DD-ME2 (P))}$ & 0.39800  & 3.38  & 3.41 & 2.22E+01 &          & 1.30   \\
				&           &       & QRPA$_{ \beta_2 (DD-PC1 (O))}$ & -0.34900 & 2.25  & 2.84 & 1.80E+01 &          & 1.05   \\
				&           &       & QRPA$_{ \beta_2 (DD-PC1 (P))}$ & 0.39700  & 3.40  & 3.41 & 2.21E+01 &          & 1.29   \\
				&           &       &       &          &       &      &          &          &        \\
				$^{74}$Kr & $\beta^+$ & 2.96  & QRPA$_{ \beta_2 (FRDM)}$  & 0.40100  & 1.59  & 1.55 & 7.67E+02 & 6.90E+02 ($\pm$ 6.6)& 1.11   \\
				&           &       & QRPA$_{ \beta_2 (NNDC)}$  & 0.25440  & 1.66  & 1.85 & 8.88E+02 &          & 1.29   \\
				&           &       & QRPA$_{ \beta_2 (IBM-1 (S))}$ & 0.00000  & 2.67  & 1.84 & 1.21E+03 &          & 1.76   \\
				&           &       & QRPA$_{ \beta_2 (DD-ME2 (O))}$ & -0.34200 & 1.77  & 1.62 & 6.92E+02 &          & 1.00   \\
				&           &       & QRPA$_{ \beta_2 (DD-ME2 (P))}$ & 0.47000  & 1.30  & 1.58 & 8.77E+02 &          & 1.27   \\
				&           &       & QRPA$_{ \beta_2 (DD-PC1 (O))}$ & -0.32500 & 1.89  & 1.57 & 6.37E+02 &          & 1.08   \\
				&           &       & QRPA$_{ \beta_2 (DD-PC1 (P))}$ & 0.48300  & 1.28  & 1.58 & 9.04E+02 &          & 1.31   \\
				&           &       &       &          &       &      &          &          &        \\
				$^{74}$Rb & $\beta^+$ & 10.42 & QRPA$_{ \beta_2 (FRDM)}$ & 0.36600  & 2.22  & 2.16 & 1.22E-01 & 6.48E-02 ($\pm$ 0.00003)& 1.88   \\
				&           &       & QRPA$_{ \beta_2 (IBM-1 (S))}$      & 0.00000  & 2.37  & 2.20 & 1.24E-01 &          & 1.92   \\
				&           &       & QRPA$_{ \beta_2 (DD-ME2 (O))}$ & -0.35600 & 2.49  & 5.22 & 2.28E-01 &          & 3.52 \\
				&           &       & QRPA$_{ \beta_2 (DD-ME2 (P))}$ & 0.45100  & 1.39  & 2.64 & 2.42E-01 &          & 3.73 \\
				&           &       & QRPA$_{ \beta_2 (DD-PC1 (O))}$ & -0.32500 & 1.90  & 5.55 & 3.64E-01 &          & 5.62  \\
				&           &       & QRPA$_{ \beta_2 (DD-PC1 (P))}$ & 0.46300  & 1.37  & 2.81 & 2.71E-01 &          & 4.18 \\
				&           &       &       &          &       &      &          &          &        \\
				$^{74}$Sr & $\beta^+$ & 10.78 & QRPA$_{ \beta_2 (FRDM)}$  & 0.40100  & 18.61 & 4.94 & 3.23E-02 & 2.76E-02 ($\pm$ 0.0026)& 1.17   \\
				&           &       & QRPA$_{ \beta_2 (IBM-1 (S))}$ & 0.00000  & 18.60 & 5.64 & 2.79E-02 &          & 1.01   \\
				&           &       & QRPA$_{ \beta_2 (DD-ME2 (O))}$ & -0.34700 & 14.67 & 5.41 & 4.36E-02 &          & 1.58   \\
				&           &       & QRPA$_{ \beta_2 (DD-ME2 (P))}$ & 0.46900  & 17.06 & 4.84 & 2.91E-02 &          & 1.06   \\
				&           &       & QRPA$_{ \beta_2 (DD-PC1 (O))}$ & -0.15800 & 17.13 & 5.21 & 2.88E-02 &          & 1.05   \\
				&           &       & QRPA$_{ \beta_2 (DD-PC1 (P))}$ & 0.48100  & 16.88 & 4.88 & 2.90E-02 &          & 1.05 \\\\
				\hline
			\end{tabular}
			%\end{adjustbox}
		\end{table*}
		
		\begin{table}[]
			\centering
			\caption{Predictive power of various pn-QRPA models used in the current investigation. See Eq.~(\ref{Rbar}) for definition of $\bar{R}$. Only four nuclei were available to calculate the $\bar{R}$ value using QRPA$_{ \beta_2 (NNDC)}$  because of limited $\beta_{2}$ values from the NNDC database.}\label{Tab12}
			\setlength{\tabcolsep}{6.5pt}
			\begin{tabular}{c|c}
				Models          & $\bar{R}$    \\
				\hline
				QRPA$_{ \beta_2 (FRDM)}$            &  1.53   \\
				QRPA$_{ \beta_2 (NNDC)}$            &  1.24   \\
				QRPA$_{ \beta_2 (IBM-1 (S))}$           &  1.89   \\
				QRPA$_{ \beta_2 (DD-ME2 (O))}$			 &  1.52 \\
				QRPA$_{ \beta_2 (DD-ME2 (P))}$			 &  1.77 \\
				QRPA$_{ \beta_2 (DD-PC1 (O))}$ 			 &  1.72   \\
				QRPA$_{ \beta_2 (DD-PC1 (P))}$			 &  1.90 \\
				\hline
			\end{tabular}
		\end{table}

		%%%%%%%%%%%%%%%%%%%%%%%%%%%%
		
		%%%%%%%%%%%%%%%%%%%%%%%%%%%%%%%%%%%%%%%%%%%%

\end{document}